\documentclass[aps,groupedaddress,preprint,superscriptaddress,showpacs]{revtex4-1}
\usepackage{eurosym}
\usepackage{amsmath}
\usepackage{fixmath}
\usepackage{graphicx}
\usepackage{wrapfig}
\usepackage[toc,page]{appendix}
\usepackage{subcaption}
\usepackage{hyperref}
\usepackage[capitalise]{cleveref}
\usepackage{subcaption}
\usepackage{amsfonts}

\graphicspath{{images/}}

\newcommand{\be}{\begin{equation}}
\newcommand{\ee}{\end{equation}}

\begin{document}

\title{The universal model of strong coupling at the nonlinear parametric resonance in open 
cavity-QED systems}

\author{ Mikhail Tokman}
\affiliation{Institute of Applied Physics, Russian Academy of Sciences, Nizhny Novgorod, 603950, Russia }
\author{ Maria Erukhimova}
\affiliation{Institute of Applied Physics, Russian Academy of Sciences, Nizhny Novgorod, 603950, Russia }
\author{Qianfan Chen}
\affiliation{Department of Physics and Astronomy, Texas A\&M University, College Station, TX, 77843 USA}
\author{Alexey Belyanin}
\affiliation{Department of Physics and Astronomy, Texas A\&M University, College Station, TX, 77843 USA}

\begin{abstract}

Many molecular, quantum-dot, and optomechanical nanocavity-QED systems demonstrate strong nonlinear interactions between electrons, photons, and phonon (vibrational) modes. We show that such systems can be described by a universal model in the vicinity of the nonlinear parametric resonance involving all three degrees of freedom. We solve the nonperturbative quantum dynamics in the strong coupling regime of the nonlinear resonance, taking into account quantization, dissipation, and fluctuations of all fields. We find analytic solutions for quantum states in the rotating wave approximation which demonstrate tripartite quantum entanglement once the strong coupling regime is reached. We show how the strong coupling at the nonlinear resonance modifies photon emission and vibrational spectra, and how the observed spectra can be used to extract information about relaxation rates and the nonlinear coupling strength in specific systems.

\end{abstract}

\date{\today }

\maketitle

\section{Introduction}

Nonlinear optical interactions acquire qualitatively new features in the strong-coupling regime of cavity quantum electrodynamics (QED), especially when utilizing an extreme field localization achievable in nanophotonic cavities. Even in the standard cavity QED scenario of the strong coupling at the two-wave resonance between only two degrees of freedom, e.g., between an electronic or vibrational transition in a molecule and an electromagnetic (EM) cavity mode, strong coupling has been shown to modify the properties of Raman scattering, generation of harmonics, four-wave mixing, and nonlinear parametric interactions, with applications in photochemistry, quantum information, and quantum sensing; see, e.g., \cite{schmidt2016, pino2015, pino2016,ramelow2019, neuman2020, ge2021,k-wang2021, narang2021, wang2021, zasedatelev2021,pscherer2021,hughes2021}. 

The dynamics becomes more complicated but also more interesting when the strong coupling regime is realized at the {\it nonlinear} resonance between three or more degrees of freedom. We will call this nonlinear resonance ``parametric'' for lack of a better word, although  this term is often applied to nonlinear processes involving only the EM fields, in which there is no energy exchange between the fields and the medium, for example parametric down-conversion in transparent nonlinear materials \cite{LLcont}. In contrast, for all examples in this paper the energy exchange between the bosonic fields and fermionic quantum emitters is of principal importance; moreover, we are interested in the nonperturbative regime of strong coupling when the excitation of the fermionic degree of freedom is not small. This is the regime most interesting for quantum technology applications as it actively involves the fermionic ``qubit'' in the processes of writing, reading, and transferring the information encoded in a quantum state. 

In this paper we will focus on the nonlinear {\it three-wave} resonance between three degrees of freedom for the sake of simpler algebra, although the formalism does not have this limitation. One possible 
example where it can be realized is a molecule or an ensemble of molecules placed in a photonic or plasmonic 
nanocavity; e.g.  \cite{benz2016, park2016,chikkaraddy2016,pscherer2021,hughes2021}. In this case the fermion
system may comprise two or more electron states forming an optical transition at frequency $\omega_e$, and the nonlinear parametric process may involve, e.g., a decay of the electron excitation into a cavity photon at frequency $\omega$ and a phonon of a given vibrational mode of a molecule at frequency $\Omega$, under the nonlinear resonance condition $\omega_e = \omega + \Omega$, or an absorption of a photon with simultaneous creation of electron and phonon excitations, given by the nonlinear resonance condition $\omega = \omega_e + \Omega$. When the strength of such a nonlinear three-wave interaction is higher than the dissipation rates, hybrid electron-photon-phonon states are formed. If the phonon mode is classical, the parametric process is simply the modulation of the electron-photon coupling by molecular vibrations which serve as an external driving force for the electron-photon quantum dynamics. If the phonon mode is quantized, the strong coupling between photon, phonon, and electron degrees of freedom near parametric resonance $\omega_e = \omega \pm \Omega$ leads inevitably to the formation of tripartite entangled states belonging to the family of Greenberger-Horne-Zeilinger (GHZ) states  \cite{tokman2021}. 

Another route to the parametric resonance is within 
the framework of  cavity optomechanics \cite{aspelmeyer2014,meystre2013,roelli2016,pirkkalainen2015} and quantum acoustics \cite{chu2017,hong2017, arriola2019}. It can occur in the situations where mechanical oscillations of a cavity parameter at frequency $\Omega$ modulate the resonance between the optical transition in a quantum emitter and a photon cavity mode. Here again the parametric resonance $\omega_e = \omega \pm \Omega$ at strong coupling should give rise to tripartite entangled states of the electrons, photons, and mechanical vibrations \cite{liao2018}. While quantization of all three degrees of freedom in experiment remains an unsolved challenge,  strong coupling and entanglement of acoustic phonons \cite{satzinger2018, bienfait2020}, resolving the energy levels of a nanomechanical oscillator \cite{arriola2019}, or cooling a macroscopic system
into its motional ground state \cite{delic2020} have already been demonstrated. 

Yet another situation leading to parametric resonance is when a quantum emitter such as a quantum dot or an optically active defect in a solid matrix is coupled to  the EM cavity field and the phonon modes of a crystal lattice \cite{wilson2002}. In this case the coupling to phonons can be introduced via the same Huang-Rhys theory as in the molecular systems \cite{weiler2012,roy2015,denning2020}; see the Hamiltonian Eq.~(\ref{mol h}) below. The exciton-photon coupling strength (Rabi frequency) in nanocavities can be even high enough to exceed the phonon frequency, which brings the system to the ultrastrong coupling regime \cite{denning2020}.

There is a great variety of models and formalisms describing these diverse physical systems, and attempts have been made to establish connections between different models. For example, it was shown in \cite{roelli2016} that plasmon-enhanced Raman scattering on individual molecules can be mapped onto a cavity optomechanics Hamiltonian when the plasmon frequency is far detuned from the electronic transition in a molecule, i.e., no electrons are excited. In this case the vibrational mode of a molecule is analogous to the mechanical oscillations of a cavity parameter. In \cite{pino2016} it was argued that Stokes Raman scattering in molecules under the condition of a strong coupling between the vibrational mode and the cavity mode is equivalent to the parametric decay of the pump photon into the Stokes photon and the vibrational quantum. In both cases a simple linear resonance $\omega = \Omega$ was assumed and the EM field was far detuned from any electronic transition in a molecule, thus excluding any real electron excitation. In \cite{hughes2021} resonant Raman scattering of single molecules when the two-wave exciton-photon coupling frequency is comparable to the vibrational frequency was analyzed. The situation when the Rabi frequency becomes comparable to the vibrational frequency was also considered in \cite{tokman2021}.  

In this paper we deal with the strong coupling at the nonlinear parametric resonance  $\omega_e = \omega \pm \Omega$ when the excitation of the electron transition and energy exchange between all three degrees of freedom are of principal importance. We show that in the RWA (i.e., excluding ultrastrong-coupling regimes)  all physical models of electron-photon-vibrational coupling in molecular, optomechanical, and any other coupled three-mode system can be mapped onto the universal ``parametric'' Hamiltonian, independently on the specific physical mechanism of coupling. Note that the system {\it must} be in the RWA regime for the nonlinear parametric resonance and all associated physics to exist and make sense. Otherwise the three-wave and two-way resonances overlap \cite{tokman2021} and the GHZ-like entangled states cannot be created. Also, it becomes  impossible to build a universal Hamiltonian. That is why the ultrastrong-coupling regimes are not of interest to us in this paper. 

When solving for the quantum dynamics of the resulting nonlinear coupled system, we include the effects of decoherence and coupling of each dynamic subsystem (electrons, photons, and phonons) to its own reservoir in the Markov approximation. Previous works (see, e.g., \cite{roy2015,denning2020}) included the non-Markovian effects in the coupling of the two-wave exciton-photon resonance to the phonon reservoir. In our case the phonon mode, which is strongly coupled to the exciton and photon modes through the nonlinear resonance, is part of the dynamical system. One could say that the phonon effect on the dynamics is ``extremely non-Markovian'', except that this terminology ceases to have any meaning in this case. The Markov approximation is of course related only to (weak) coupling of all components of the dynamic system to their dissipative reservoirs. 

Within the formalism of the stochastic equation of evolution for the state vector we are able to find the general analytic solution for the nonperturbative dynamics of the open quantum system. This approach is well known \cite{zoller1997}, but it is usually applied for numerical Monte-Carlo simulations \cite{scully1997,plenio1998,gisin1992,diosi1998,tannoudji1993,molmer1993,gisin1992-2,raedt2017,nathan2020}. We recently developed a version of stochastic Schr\"{o}dinger equation suitable for analytic solutions in open strongly-coupled cavity QED problems  \cite{tokman2021, chen2021}. We calculate the photon and phonon emission spectra to obtain the experimentally observable signatures of the strong coupling regime and tripartite quantum entanglement. It is remarkable that we are able to find analytic solutions for the quantum dynamics in systems of coupled electron, photon, and phonon excitations even including dissipation and fluctuation effects in all subsystems. This allows one to find the expressions for the time evolution of the state vector and observable quantities in the form which shows explicitly the dependence on all experimental parameters: transition energies and frequencies, matrix elements of the optical transitions, the spatial structure of the field modes, relaxation rates for all constituent subsystems, ambient temperatures etc. We believe that the results obtained in this paper will be useful for designing and interpreting the experiments on a broad range of cavity QED systems.

The paper is structured as follows. In Section II we present the Hamiltonian for coupled quantized fermion, photon, and phonon fields near the parametric resonance for one particular mechanism of three-wave coupling. In Section III we show that a large variety of different three-wave coupling mechanisms and physical systems are reduced to the same Hamiltonian which therefore can serve as a universal model of the parametric resonance.  Section IV includes the effects of dissipation, decoherence, and fluctuations within the stochastic equation for the state vector which describes the evolution of  an open system in contact with dissipative reservoirs.  As compared to our recent work  \cite{tokman2021, chen2021}, we develop a model of fluctuations and dissipative processes which includes all effects of phonon dissipation on the dynamics of the parametric process and the emission spectra.  In Section V we describe the formation of entangled electron-photon-phonon states for an open system. Section VI calculates the emission spectra of photons and phonons resulting from the nonlinear parametric decay of an electron excitation.  Appendix A contains the derivation details for some results in Sec. V.

\section{The universal Hamiltonian of the nonlinear parametric resonance}

Consider a simple model of three interacting quantum subsystems which includes (1)
an electron transition in a quantum emitter such as an atom, molecule,
optically active impurity, quantum dot, etc., which we will model as a 2-level system, (2) a single-mode
electromagnetic (EM) field in a cavity, and (3) a mode of mechanical,
acoustic, or molecular vibrations (``phonons''). Although in this section we write the Hamiltonian for a specific model, in the next section we show that the same Hamiltonian describes the nonlinear parametric coupling in a variety of physical systems. 

The generalization to many bosonic modes or fermionic degrees of freedom is straightforward and still allows analytic solution within the rotating wave
approximation (RWA), but it leads to more cumbersome algebra (see, e.g., \cite{tokman2021-3}), so we will keep only three degrees of freedom for clarity. 

Figure 1 shows a generic model of parametric decay of an electron excitation in a quantum emitter (e.g., a molecule) into a photon of a cavity mode at frequency $\omega$ and a phonon of a given vibrational mode  at frequency $\Omega$, under the condition of the parametric resonance $\omega_e \approx \omega + \Omega$.  

%%%%%%%%%%%%%%%%%%%%%%%%%%%%%%%%%% 

\begin{figure}[htb]
\includegraphics[width=0.4\linewidth]{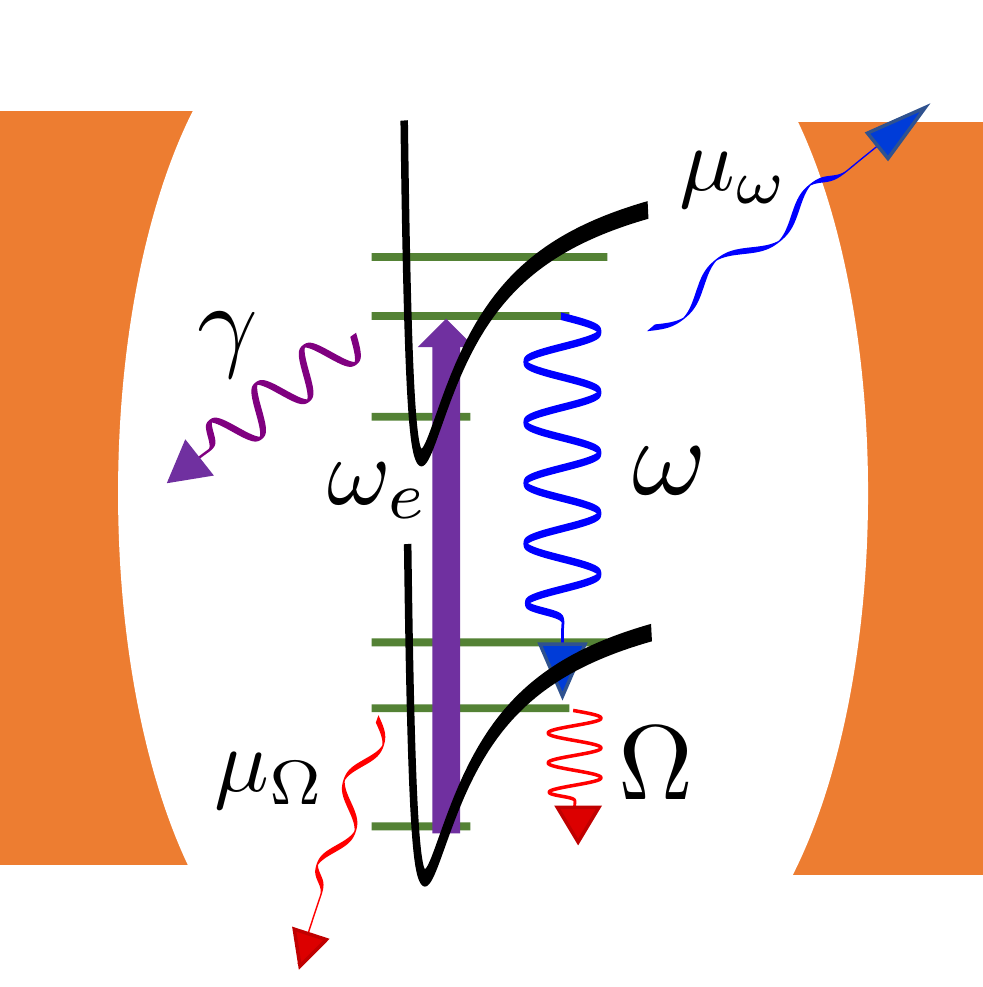}
\caption{ A sketch of nonlinear parametric resonance for a molecule in a cavity  showing the decay of the electron excitation at frequency $\omega_e$ into a cavity mode photon at frequency $\omega$ and a phonon of a given vibrational mode  at frequency $\Omega$. The relaxation rates of the electron, photon, and vibrational excitations are $\gamma$, $\mu_{\omega}$, and $\mu_{\Omega}$, respectively. }
\label{fig1}
\end{figure}

In the absence of coupling, the partial Hamiltonians are:

%%%%%%%%%%%%%%%%%%%%%%%
\subsection{The 2-level fermion system}
It is described by a standard effective Hamiltonian
\begin{equation}
\hat{H}_{e}=\hbar \omega_e\hat{\sigma}^{\dagger }\hat{\sigma}.  \label{ha}
\end{equation}
Here $\hat{\sigma}=\left\vert 0\right\rangle \left\langle 1\right\vert $, $
\hat{\sigma}^{\dagger }=\left\vert 1\right\rangle \left\langle 0\right\vert $
; $\left\vert 0\right\rangle $ and $\left\vert 1\right\rangle $ are the
eigenstates of an ``atom'' with energies $0$ and $\hbar \omega_e$
respectively. The Hamiltonian (\ref{ha}) corresponds to the dipole moment
operator
\begin{equation}
\mathbf{\hat{d}}=\mathbf{d}\left( \hat{\sigma}^{\dagger }+\hat{\sigma}%
\right) ,  \label{dip}
\end{equation}%
where $\mathbf{d=-e}\left\langle 1\right\vert \mathbf{r}\left\vert
0\right\rangle $, $\mathbf{r}$ is a coordinate for the finite motion of a bound 
electron.
%%%%%%%%%%%%%%%%%%%%%%%%%%

\subsection{The EM field}

Here we consider a single-mode EM field for simplicity, although including many bosonic field modes does not present any principal difficulties. Besides, in a micro- or nanocavity other EM modes will be far detuned from the nonlinear resonance. The Hamiltonian is 
\begin{equation}
\hat{H}_{em}=\hbar \omega \hat{c}^{\dagger }\hat{c}.  \label{hem}
\end{equation}%
Here $\hat{c}$ and $\hat{c}^{\dagger }$ are standard bosonic annihilation
and creation operators of photons or plasmons in the EM mode of frequency $%
\omega $ . The electric field operator is
\begin{equation}
\mathbf{\hat{E}}=\mathbf{E}\left( \mathbf{r}\right) \hat{c}+\mathbf{E}^{\ast
}\left( \mathbf{r}\right) \hat{c}^{\dagger }.  \label{ef}
\end{equation}%
The spatial structure of the normalization amplitude of the field $\mathbf{E}%
\left( \mathbf{r}\right) $ is determined by solving the boundary value
problem. The normalization condition is
\begin{equation}
\int_{V}\frac{\partial \left[ \omega ^{2}\varepsilon \left( \omega ,\mathbf{r%
}\right) \right] }{\omega \partial \omega }\mathbf{E}^{\ast }\left( \mathbf{r%
}\right) \mathbf{E}\left( \mathbf{r}\right) d^{3}r=4\pi \hbar \omega .
\label{nc}
\end{equation}

Here $V$ is the quantization volume, $\varepsilon \left( \omega ,\mathbf{r}%
\right) $ the dielectric function of the dispersive medium which fills in
the resonator. Eq.~(\ref{nc}) is derived, e.g., in \cite%
{tokman2016,tokman2013,tokman2015,tokman2018}.
%%%%%%%%%%%%%%%%%%%%%%%%%%%%%%%%%%

\subsection{The phonons}

We again assume a single bosonic mode of a vibrational field for the same reasons, 
\begin{equation}
\hat{H}_{p}=\hbar \Omega \hat{b}^{\dagger }\hat{b},  \label{hp}
\end{equation}%
where $\hat{b}$ and $\hat{b}^{\dagger }$ are phonon annihilation and
creation operators. Depending on the situation, they may define, e.g., the 
radius-vector of oscillations of the center of mass of an atom \cite%
{tokman2021,neuman2020,felipe2017} or a geometric parameter of the
optomechanical cavity \cite{roelli2016,aspelmeyer2014},
\begin{equation}
\mathbf{\hat{R}}=\mathbf{Q}\hat{b}+\mathbf{Q}^{\ast }\hat{b}^{\dagger }.
\label{af}
\end{equation}%
The normalization amplitude $\mathbf{Q}$ depends on the system; its absolute
value can be expressed through an effective mass of the quantum mechanical
oscillator \cite{aspelmeyer2014}: $\left\vert \mathbf{Q}\right\vert ^{2}=%
\frac{\hbar }{2m_{eff}\Omega }$. 
%%%%%%%%%%%%%%%%%%%%%%%%%%%%%%

\subsection{The coupling}

The coupling between subsystems is
strongest at resonance. Since usually $\omega ,\omega_e\gg \Omega $, two most relevant resonances are a two-wave resonance,
\begin{equation}
\omega_e\approx \omega  \label{two-wave resonance}
\end{equation}%
and a three-wave (parametric) resonance,
\begin{equation}
\omega_e\approx \omega \pm \Omega
\label{three-wave (parametric) resonance}
\end{equation}%
There could be also harmonic resonances at the harmonics of the phonon
frequency: $\omega_e\approx \omega \pm M\Omega $, where $M$ is integer.
The modulation of the system parameters by a classical phonon field at
frequency $\Omega $ was studied, e.g., in \cite{chen2021}, and we
don't consider the classical field here.

The two-wave resonance in the RWA \cite{scully1997} is desrcibed by the Jaynes-Cummings (JC) 
 Hamiltonian \cite{jaynes1963}, 
\begin{equation}
\hat{H}=\hbar \omega \hat{c}^{\dagger }\hat{c}+\hbar \omega_e\hat{\sigma}%
^{\dagger }\hat{\sigma}+\hbar \left( \Omega _{R}^{\left( 2\right) }\hat{%
\sigma}^{\dagger }\hat{c}+h.c.\right) .  \label{twrh}
\end{equation}%
The electric dipole coupling between the electron and EM subsystems is
expressed here through the effective Rabi frequency for the two-wave
coupling, $\Omega _{R}^{\left( 2\right) }=$ $-\frac{\mathbf{d}\cdot \mathbf{E%
}\left( \mathbf{r}_{0}\right) }{\hbar }$, where $\mathbf{r}_{0}$ is the
coordinate of a point-like atom.

A three-wave (parametric) resonance appears in many different scenarios. As
we show in the next section, various models existing in the literature can
be described with one universal Hamiltonian. One of the scenarios leading to
the universal three-wave coupling Hamiltonian is when a quantized phonon
(vibrational) mode modulates the coupling energy between the
 electron transition and the EM
field mode. In this case the JC Hamiltonian is generalized to the following form \cite%
{tokman2021,pino2016}:
\begin{equation}
\hat{H}=\hbar \omega \hat{c}^{\dagger }\hat{c}+\hbar \omega_e\hat{\sigma}%
^{\dagger }\hat{\sigma}+\hbar \Omega \hat{b}^{\dagger }\hat{b}+\hbar \left(
\Omega _{R}^{\left( 3\right) }\hat{\sigma}^{\dagger }\hat{c}\hat{b}%
+h.c.\right) ,  \label{prh}
\end{equation}%
where $\Omega _{R}^{\left( 3\right) }$ is the coupling parameter for the
three-wave resonance, which depends on the specific coupling mechanism. The above interaction term is written for the decay of an
electron transition into the photon and the phonon, i.e., assuming that the
electron excitation energy is the largest of the three. This decay process
corresponds to the upper (plus) sign in the resonant condition (\ref%
{three-wave (parametric) resonance}). If we choose the lower (minus) sign in Eq.~(\ref{three-wave
(parametric) resonance}), the three-wave coupling Hamiltonian will become  $\hbar
\left( \Omega _{R}^{\left( 3\right) }\hat{c}^{\dagger }\hat{b}\hat{\sigma}%
+h.c.\right) $.

Note that the
three-wave coupling term in Eq.~(\ref{prh}) has the structure formally equivalent to the
parametric down-conversion (PDC) Hamiltonian describing the parametric decay
of the quantized pump field into quantized signal and idler modes \cite{couteau2018,tokman2021-2}. Of
course one difference is that all fields in the photonic PDC process are
described by bosonic operators whereas the electron excitation in Eq.~(\ref%
{prh}) is described by fermionic operators, giving rise to its specific nonlinearities.

The strong coupling regime is realized when the three-wave coupling
parameter in Eq.~(\ref{prh}) is larger than a certain combination of the relaxation constants $
\gamma, \mu_{\omega}, \mu_{\Omega}$ of all subsystems. The exact criterion can be retrieved from the analytic solution presented in Sections V and VI below. 

The specific form of the parameter $\Omega _{R}^{\left( 3\right) }$ depends
on the nonlinear coupling mechanism. For example, a phonon
mode can modulate the position of the center of mass of an ``atom'' within a spatially nonuniform distribution of the EM
field of a cavity mode. It can be realized for all kinds of quantum
emitters: an electron transition in a molecule, a quantum dot or defect in a
solid matrix, an optomechanical system with a varying cavity parameter, etc. In this case, in the limit of a small amplitude of
vibrations, one can obtain that \cite{tokman2021}
\begin{equation}
\Omega _{R}^{\left( 3\right) }=-\frac{1}{\hbar }\left[ \mathbf{d}\left(
\mathbf{Q\cdot \nabla }\right) \mathbf{E}\right] _{\mathbf{r}=\mathbf{r}%
_{0}}.  \label{omega3}
\end{equation}%
The Hamiltonian in Eq.~(\ref{prh}) is valid if the three-wave resonance is
well separated from the two-wave one. The conditions for that are \cite{tokman2021}
\begin{equation}
\left\vert \omega_e-\omega -\Omega \right\vert \ll \left\vert \omega
_{e}-\omega \right\vert ,\ \left\vert \Omega _{R}^{\left( 2,3\right)
}\right\vert \ll \Omega .  \label{condition}
\end{equation}

In the next section we will see that if the conditions (\ref{condition}) are
satisfied, other models of three-wave coupling can be reduced to the universal 
parametric Hamiltonian (\ref{prh}). Note that in plasmonic nanocavities the two-wave or/and three-wave Rabi frequency $
\Omega_R^{(2,3)}$ can become higher than the vibrational or phonon frequency $\Omega$, which would violate the last of inequalities (\ref{condition}); see \cite{hughes2021,denning2020,tokman2021}. 

%%%%%%%%%%%%%%%%%%%%%%%%%%%%%%%%%%%

\section{The models described by the universal parametric Hamiltonian }

In addition to the three-wave coupling mechanism considered in the previous
section, there are other ways for phonons or any mechanical oscillations to affect the
coupling between the EM cavity field and the quantum emitter. Here we give
several examples, assuming without loss of generality that the amplitudes $%
\mathbf{Q}$ in the expression for the position displacement operator in Eq.~(%
\ref{af}) are real functions.
%%%%%%%%%%%%%%%%%%%%%%%%%%%%

\subsection{Phonons modulate the energy of the electron transition (see, e.g.,
\cite{neuman2020,felipe2017})}

\begin{equation}
\hat{H}=\hbar \omega \hat{c}^{\dagger }\hat{c}+\hbar \omega_e\hat{\sigma}%
^{\dagger }\hat{\sigma}+\hbar \Omega \hat{b}^{\dagger }\hat{b}+\hbar \left(
\Omega _{R}^{\left( 2\right) }\hat{\sigma}^{\dagger }\hat{c}+h.c.\right)
+\hbar \sqrt{S}\Omega \hat{\sigma}^{\dagger }\hat{\sigma}\left( \hat{b}+\hat{%
b}^{\dagger }\right) .  \label{mol h}
\end{equation}%
Here $S$ is the Huang--Rhys factor, which determines the dependence of
the transition energy on the dimensionless amplitude $\hat{b}+\hat{b}%
^{\dagger }$ of the phonon oscillations. Let's call Eq.~(\ref{mol h}) the
``molecular'' Hamiltonian, although it can also describe the exciton-phonon coupling in quantum-dot systems \cite{weiler2012,roy2015,denning2020}. 

%%%%%%%%%%%%%%%%%%%%%%%

\subsection{Phonons modulate some geometric parameter of the 
cavity (see, e.g., \cite{roelli2016,aspelmeyer2014})}

\begin{equation}
\hat{H}=\hbar \omega \hat{c}^{\dagger }\hat{c}+\hbar \omega_e\hat{\sigma}%
^{\dagger }\hat{\sigma}+\hbar \Omega \hat{b}^{\dagger }\hat{b}+\hbar \left(
\Omega _{R}^{\left( 2\right) }\hat{\sigma}^{\dagger }\hat{c}+h.c.\right)
-\hbar g\hat{c}^{\dagger }\hat{c}\left( \hat{b}+\hat{b}^{\dagger }\right) .
\label{opmech h}
\end{equation}%
Here the factor $g$ determines the linear dependence of the resonant
frequency of the cavity on some geometric parameter $\mathbf{G}$ modulated
by mechanical vibrations:%
\begin{equation*}
g=-\mathbf{Q}\frac{\partial \omega }{\partial \mathbf{G}}.
\end{equation*}%
We will call Eq.~(\ref{opmech h}) the ``optomechanical'' Hamiltonian.

The Hamiltonians (\ref{mol h}) and (\ref{opmech h}) appear to be very different from Eq.~(\ref{prh}). Indeed, they both contain the standard two-wave resonance and their three-wave coupling terms are different. We will show now that when the conditions (\ref{condition}) are satisfied,
the ``molecular'' and ``optomechanical'' Hamiltonians are equivalent to the
universal Hamiltonian in Eq.~(\ref{prh}).
To prove this statement, we write the Hamiltonian (\ref{prh}) in the
interaction representation:%
\begin{equation}
\hat{H}_{int}=e^{i\hat{H}_{0}t}\hat{V}e^{-i\hat{H}_{0}t},  \label{int pic}
\end{equation}%
where
\begin{equation*}
\hat{H}_{0}=\hbar \omega \hat{c}^{\dagger }\hat{c}+\hbar \omega_e\hat{%
\sigma}^{\dagger }\hat{\sigma}+\hbar \Omega \hat{b}^{\dagger }\hat{b},\ \
\hat{V}=\hbar \left( \Omega _{R}^{\left( 3\right) }\hat{\sigma}^{\dagger }%
\hat{c}\hat{b}+h.c.\right) .
\end{equation*}%
This yields

\begin{equation}
\hat{H}_{int}=\hbar \left( \Omega _{R}^{\left( 3\right) }\hat{\sigma}%
^{\dagger }\hat{c}\hat{b}e^{i\left( \omega_e-\omega -\Omega \right)
t}+h.c.\right) .  \label{iph}
\end{equation}

Next, we write the Hamiltonians in Eqs.~(\ref{mol h}) and (\ref{opmech h})
in the interaction representation by defining the unperturbed Hamiltonian as
\begin{equation}
\hat{H}_{0}=\hbar \omega \hat{c}^{\dagger }\hat{c}+\hbar \omega_e\hat{%
\sigma}^{\dagger }\hat{\sigma}+\hbar \Omega \hat{b}^{\dagger }\hat{b}+\hbar
\left( \Omega _{R}^{\left( 2\right) }\hat{\sigma}^{\dagger }\hat{c}%
+h.c.\right) .  \label{uph}
\end{equation}%
This gives%
\begin{equation}
\hat{V}=\hat{V}_{mol}=\hbar \sqrt{S}\Omega \hat{\sigma}^{\dagger }\hat{\sigma%
}\left( \hat{b}+\hat{b}^{\dagger }\right)  \label{vmol}
\end{equation}%
for the ``molecular'' Hamiltonian and
\begin{equation}
\hat{V}=\hat{V}_{optom}=-\hbar g\hat{c}^{\dagger }\hat{c}\left( \hat{b}+\hat{%
b}^{\dagger }\right)  \label{voptom}
\end{equation}%
for the optomechanical one.

The operator $\hat{H}_{0}$ given by Eq.~(\ref{uph}) is not diagonal. In this
case one should generally diagonalize $\hat{H}_{0}$. However, sometimes a
slightly different approach is simpler. Indeed, consider the Hamiltonian
given by%
\begin{equation}
\hat{H}=\hat{H}_{0}\left( \hat{A}_{1,}\cdots \hat{A}_{N}\right) +\hat{V}%
\left( \hat{A}_{1,}\cdots \hat{A}_{N}\right) ,  \label{Hamiltonian}
\end{equation}%
where $\hat{A}_{i}$ are certain operators related to coupled subsystems
(here we assume that Hermitian conjugated operators are assigned different
numbers ``$i$'' ). For any interaction
operator $\hat{V}$, which can be expanded in a series%
\begin{equation*}
\hat{V}=\sum_{j}k_{j}\prod_{i=1}^{N_{j}}\left( \hat{A}_{i}\right) ^{n_{ji}}
\end{equation*}%
(here $n_{ji}$ are positive integers), the following relationships are true:
\begin{eqnarray*}
\hat{H}_{int} &=&e^{i\hat{H}_{0}t}\hat{V}\left( \hat{A}_{1,}\cdots \hat{A}%
_{N}\right) e^{-i\hat{H}_{0}t} \\
&=&\sum_{j}k_{j}\prod_{i=1}^{N_{j}}e^{i\hat{H}_{0}t}\left( \hat{A}%
_{i}\right) ^{n_{ji}}e^{-i\hat{H}_{0}t} \\
&=&\sum_{j}k_{j}\prod_{i=1}^{N_{j}}\left( e^{i\hat{H}_{0}t}\hat{A}_{i}e^{-i%
\hat{H}_{0}t}\right) ^{n_{ji}},
\end{eqnarray*}%
from which one obtains%
\begin{equation}
\hat{H}_{int}=\hat{V}\left( \widehat{\tilde{A}}_{1,}\cdots \widehat{\tilde{A}%
}_{N,}\right) ,  \label{int h}
\end{equation}%
where%
\begin{equation}
\widehat{\tilde{A}}_{i}\left( t,\hat{A}_{1,}\cdots \hat{A}_{N}\right) =e^{i%
\hat{H}_{0}t}\hat{A}_{i}e^{-i\hat{H}_{0}t}  \label{Ai}
\end{equation}%
The operators $\widehat{\tilde{A}}_{i}$ satisfy the Heisenberg equations
\begin{equation}
\frac{\partial \widehat{\tilde{A}}_{i}}{\partial t}=\frac{i}{\hbar }\left[
\hat{H}_{0},\widehat{\tilde{A}}_{i}\right]   \label{he}
\end{equation}%
for the initial conditions $\widehat{\tilde{A}}_{i}\left( t=0\right) =\hat{A}%
_{i}$. In particular, for the Hamiltonian $\hat{H}_{0}$ given by Eq.~(\ref%
{uph}) one obtains
\begin{equation}
\frac{\partial \widehat{\tilde{\sigma}}}{\partial t}=-i\omega_e\widehat{%
\tilde{\sigma}}-i\Omega _{R}^{\left( 2\right) }\widehat{\tilde{c}}\left( 1-2%
\widehat{\tilde{\sigma}}^{\dagger }\widehat{\tilde{\sigma}}\right) ,
\label{he1}
\end{equation}%
\begin{equation}
\frac{\partial \widehat{\tilde{c}}}{\partial t}=-i\omega \widehat{\tilde{c}}%
-i\Omega _{R}^{\left( 2\right) \ast }\widehat{\tilde{\sigma}},  \label{he2}
\end{equation}%
where we used the integral of motion $\widehat{\tilde{\sigma}}^{\dagger }%
\widehat{\tilde{\sigma}}+\widehat{\tilde{\sigma}}\widehat{\tilde{\sigma}}%
^{\dagger }=1$. Taking into account the condition $\left\vert \Omega
_{R}^{\left( 2\right) }\right\vert \ll \left\vert \omega_e-\omega
\right\vert $ which follows from Eqs. (\ref{condition}), when solving for
the operators $\widehat{\tilde{c}}$ and $\widehat{\tilde{\sigma}}$ one can
neglect the terms of the order of $\left\vert \frac{\Omega _{R}^{\left(
2\right) }}{\omega_e-\omega }\right\vert ^{2}$. For the initial
conditions $\widehat{\tilde{\sigma}}$ $\left( t=0\right) =\hat{\sigma}$ and $%
\widehat{\tilde{c}}\left( t=0\right) =\hat{c}$ the solution expanded in
series in powers of the small parameter $\left\vert \frac{\Omega
_{R}^{\left( 2\right) }}{\omega_e-\omega }\right\vert $ takes the form
\begin{equation}
\left(
\begin{array}{c}
\widehat{\tilde{\sigma}} \\
\widehat{\tilde{c}}%
\end{array}%
\right) =\widehat{\widehat{M}}\left(
\begin{array}{c}
\hat{\sigma} \\
\hat{c}%
\end{array}%
\right) ,  \label{sol}
\end{equation}%
where%
\begin{eqnarray}
\widehat{\widehat{M}} &=&\left(
\begin{array}{cc}
e^{-i\omega_et} & 0 \\
0 & e^{-i\omega t}%
\end{array}%
\right)   \notag \\
&&-\left(
\begin{array}{cc}
0 & \frac{\Omega _{R}^{\left( 2\right) }}{\omega_e-\omega }\left( 1-2\hat{%
\sigma}^{\dagger }\hat{\sigma}\right) \left( e^{-i\omega t}-e^{-i\omega
_{e}t}\right)  \\
\frac{\Omega _{R}^{\left( 2\right) \ast }}{\omega_e-\omega }\left(
e^{-i\omega t}-e^{-i\omega_et}\right)  & 0%
\end{array}%
\right) +o\left( \left\vert \frac{\Omega _{R}^{\left( 2\right) }}{\omega
_{e}-\omega }\right\vert ^{2}\right)   \label{M}
\end{eqnarray}

There is an exact solution for the operator $\widehat{\tilde{b}}$:
\begin{equation}
\widehat{\tilde{b}}=e^{i\hat{H}_{0}t}\hat{b}e^{-i\hat{H}_{0}t}=\hat{b}%
e^{-i\Omega t}.  \label{b}
\end{equation}

Now we substitute the three-wave coupling Hamiltonian (\ref{vmol}) into Eq.~(%
\ref{int h}) and use Eqs.~(\ref{sol})-(\ref{b}). The conditions (\ref%
{condition}) combined with the RWA allow one to keep only slowly varying
terms $\propto e^{i\left( \omega_e-\omega -\Omega \right) t}$ in the
final expression. Taking into account that $\hat{\sigma}^{\dagger }\hat{%
\sigma}^{\dagger }=\hat{\sigma}\hat{\sigma}$ $=0$ and taking $\frac{1}{%
\omega_e-\omega }\approx \frac{1}{\Omega }$ in Eq.~(\ref{M}), we obtain
the following expression for the ``
molecular'' Hamiltonian in the interaction picture:%
\begin{equation}
\left( \hat{H}_{int}\right) _{mol}=-\hbar \left( \sqrt{S}\Omega _{R}^{\left(
2\right) }\hat{\sigma}^{\dagger }\hat{c}\hat{b}e^{i\left( \omega_e-\omega
-\Omega \right) t}+h.c.\right)  \label{mh int}
\end{equation}
A similar derivation for the ``
optomechanical'' Hamiltonian given by Eq.~(\ref{voptom})
leads to the following result:

\begin{equation}
\left( \hat{H}_{int}\right) _{optom}=-\hbar \left( \frac{g\Omega
_{R}^{\left( 2\right) }}{\Omega }\hat{\sigma}^{\dagger }\hat{c}\hat{b}%
e^{i\left( \omega_e-\omega -\Omega \right) t}+h.c.\right)
\label{optomh int}
\end{equation}

Clearly, in both cases the Hamiltonian has the same structure as the
parametric Hamiltonian (\ref{iph}), in which
\begin{equation}
\left( \Omega _{R}^{\left( 3\right) }\right) _{mol}=-\sqrt{S}\Omega
_{R}^{\left( 2\right) }\ \ {\rm or}\ \ \left( \Omega _{R}^{\left( 3\right)
}\right) _{optom}=-\frac{g}{\Omega }\Omega _{R}^{\left( 2\right) }.
\label{omega 3}
\end{equation}%
Therefore, one can use the universal parametric Hamiltonian (\ref{prh}) for
all kinds of three-wave couplings after choosing an appropriate expression
for the coupling parameter $\Omega _{R}^{\left( 3\right) }$. 

We emphasize again that the universal character of the parametric
Hamiltonian (\ref{prh}) holds as long as inequalities (\ref{condition}) are
satisfied, which ensure that the three-wave nonlinear resonance can be separated from
the two-wave resonance.

%%%%%%%%%%%%%%%%%%%%%%%%%%%%%%%

\section{Including dissipation and fluctuations within the
stochastic equation for the state vector}

\subsection{Stochastic equation for the state vector}

Many quantum information applications are based on the
strong coupling regime in which the relaxation times $\tau $ are much longer
than the dynamical coupling times $T$ between the subsystems. For two- and
three-wave couplings those times are determined by effective Rabi
frequencies, $T$ $^{-1}$ $\sim \left\vert \Omega _{R}^{\left( 2,3\right)
}\right\vert $. As shown in \cite{tokman2021,chen2021}, the method of the
stochastic equation for the state vector is often the most convenient way to
describe the nonperturbative dynamics of open strongly coupled systems; it leads to simpler
derivations for the observables and characterization of entanglement than
the operator-valued Heisenberg-Langevin equations or the master equation for
the density matrix.

Indeed, when applied to strongly coupled systems the Heisenberg approach
leads to the nonlinear operator-valued equations even in the simplest case
of a single two-level atom coupled to a single-mode field \cite{scully1997}.
In contrast, the equations for the state vector components are always linear; they
contain much fewer variables as compared to the density matrix equations and
are split into low-dimensional blocks in the RWA, leading to analytic
solutions for both two-wave \cite{scully1997} and three-wave \cite%
{tokman2021,chen2021} resonant coupling.

One potential difficulty with this approach is that dissipation and fluctuations may lead to the
coupling between different blocks of equations for the state vector
components that were uncoupled in a closed system. However, in the strong
coupling regime the coupling through dissipative reservoirs is weak (scales
as a small parameter $T/\tau $) and can be taken into account perturbatively
\cite{tokman2021,chen2021}.

Following \cite{tokman2021,chen2021}, we apply the stochastic equation for
the state vector to derive analytic solution for the parametric coupling of
a two-level fermionic quantum emitter to two boson fields. The stochastic equation has
the following general form, 
\begin{equation}
\frac{d}{dt}\left\vert \Psi \right\rangle =-\frac{i}{\hbar }\hat{H}%
_{eff}\left\vert \Psi \right\rangle -\frac{i}{\hbar }\left\vert \mathfrak{R}%
\right\rangle .  \label{stochastic eq for psi}
\end{equation}%
Here $\left\vert \Psi \right\rangle $ is the state vector; $\left\vert
\mathfrak{R}\right\rangle $ is the noise term satisfying $\overline{%
\left\vert \mathfrak{R}\right\rangle }=0$, where the bar means averaging over the noise
statistics; $\hat{H}_{eff}=$ $\hat{H}+\hat{H}^{\left( ah\right) }$ is an
effective Hamiltonian which is a non-Hermitian operator. Its non-Hermitian
component $\hat{H}^{\left( ah\right) }$ describes the effects of relaxation.
The expressions for $\hat{H}^{\left( ah\right) }$ and $\left\vert \mathfrak{R%
}\right\rangle $ must be consistent with each other to guarantee the
conservation of the noise-averaged norm, $\overline{\left\langle \Psi \left(
t\right) \right. \left\vert \Psi \left( t\right) \right\rangle }=1$, and
ensure that the system reaches a physically meaningful steady state in the
absence of any external driving force. To calculate the observables from the
state vector given by Eq.~(\ref{stochastic eq for psi}) one should apply a
standard procedure but with an important extra step: averaging over the
noise statistics, i.e., $q=\overline{\left\langle \Psi \right\vert \hat{q}%
\left\vert \Psi \right\rangle }$ , where $\hat{q}$ is a quantum-mechanical
operator corresponding to the observable $q$. 

Perhaps the most popular version of the stochastic approach to derive the
state vector, i.e., the stochastic Schr\"{o}dinger equation (SSE), is its
application for numerical Monte-Carlo simulations within the method of
quantum jumps \cite{scully1997,plenio1998,gisin1992,diosi1998,tannoudji1993,molmer1993,gisin1992-2,raedt2017,nathan2020}. The stochastic equation in a different form, the Schr\"{o}dinger-Langevin
equation (SLE), was suggested to describe the Brownian motion of a quantum
particle in a constant field \cite{kostin1972,katz2016}. Generally, using
some version of the stochastic equation fits within the
narrative of the Langevin method \cite{WM94}. Within the Langevin approach which
describes the system with stochastic equations of evolution, the averaging
over the reservoir degrees of freedom is equivalent to averaging over the
statistics of the noise sources \cite{landau1965}. This paradigm allows one
to describe open systems without relying on the density matrix.

%%%%%%%%%%%%%%%%%%%%%%%%%%%%%%

\subsection{Comparison with the Lindblad formalism}

It was shown in \cite{tokman2021} that one can choose the form of $\hat{H}%
^{\left( ah\right) }$ and $\left\vert \mathfrak{R}\right\rangle $ in such a
way that the observables calculated with Eq.~(\ref{stochastic eq for psi})
will coincide with those obtained by solving the master equation in the
Lindblad approximation. The corresponding master equation has the form \cite%
{scully1997}
\begin{equation}
\frac{d}{dt}\hat{\rho}=-\frac{i}{\hbar }\left[ \hat{H},\hat{\rho}\right] +%
\hat{L}\left( \hat{\rho}\right)  \label{Lind mas eq}
\end{equation}%
where $\hat{L}\left( \hat{\rho}\right) $ is the relaxation operator
(Lindbladian) which can be represented as
\begin{equation}
\hat{L}\left( \hat{\rho}\right) =-\frac{i}{\hbar }\left( \hat{H}^{\left(
ah\right) }\hat{\rho}-\hat{\rho}\hat{H}^{\left( ah\right) \dagger }\right)
+\delta \hat{L}\left( \hat{\rho}\right) .  \label{L rho}
\end{equation}%
The equivalence (in the above sense) between the stochastic equation and the
Lindblad approach exists if we substitute the anti-Hermitian part of the
Hamiltonian from Eq.~(\ref{L rho}) into Eq.~(\ref{stochastic eq for psi}),
and postulate the following correlation properties for the noise source:%
\begin{equation}
\overline{\left\vert \mathfrak{R}\left( t^{\prime }\right) \right\rangle
\left\langle \mathfrak{R}\left( t^{\prime \prime }\right) \right\vert }%
=\hbar ^{2}\delta \left( t^{\prime }-t^{\prime \prime }\right) \delta \hat{L}%
\left( \hat{\rho}\right) _{\hat{\rho}\Longrightarrow \overline{\left\vert
\Psi \right\rangle \left\langle \Psi \right\vert }}
\label{correlation properties}
\end{equation}

%%%%%%%%%%%%%%%%%%%%%%%%%%%%%%%%

\subsection{Parametric decay of the electron excitation into a photon and a
phonon}

We will describe the dynamics near the three-wave electron-photon-phonon
resonance by the stochastic equation (\ref{stochastic eq for psi}) with
the parametric Hamiltonian (\ref{prh}). We will seek the state vector in the
form
\begin{equation*}
\Psi =\sum_{n,\alpha =0}^{\infty ,\infty }\left( C_{\alpha n0}\left\vert
\alpha \right\rangle \left\vert n\right\rangle \left\vert 0\right\rangle
+C_{\alpha n1}\left\vert \alpha \right\rangle \left\vert n\right\rangle
\left\vert 1\right\rangle \right) ,
\end{equation*}%
where the order of indices corresponds to
\begin{equation*}
C_{\mathrm{phonon\,photon\,fermion}}\left\vert phonon\right\rangle
\left\vert photon\right\rangle \left\vert fermion\right\rangle .
\end{equation*}%
Consider the initial state with an excited electron and no bosonic excitations,
of the type $\left\vert \Psi \left( 0\right) \right\rangle =\left\vert
0\right\rangle \left\vert 0\right\rangle \left\vert 1\right\rangle $ . Near
the resonance $\omega_e\approx \omega +\Omega $ the three-wave coupling
leads to the excitation of the state $\left\vert 1\right\rangle \left\vert
1\right\rangle \left\vert 0\right\rangle $. In the zero-temperature limit,
which is valid when the reservoir temperature is much lower than the transition
frequency (for optical frequencies it is satisfied even at room
temperature), relaxation processes could populate only the states with lower
energies: $\left\vert 0\right\rangle \left\vert 1\right\rangle \left\vert
0\right\rangle $, $\left\vert 1\right\rangle \left\vert 0\right\rangle
\left\vert 0\right\rangle $ and $\left\vert 0\right\rangle \left\vert
0\right\rangle \left\vert 0\right\rangle $. Therefore, for this initial
condition the state vector will have 5 components:
\begin{eqnarray}
\left\vert \Psi \left( t\right) \right\rangle &=&C_{000}\left( t\right)
\left\vert 0\right\rangle \left\vert 0\right\rangle \left\vert
0\right\rangle +C_{010}\left( t\right) \left\vert 0\right\rangle \left\vert
1\right\rangle \left\vert 0\right\rangle +C_{100}\left( t\right) \left\vert
1\right\rangle \left\vert 0\right\rangle \left\vert 0\right\rangle  \notag \\
&&+C_{110}\left( t\right) \left\vert 1\right\rangle \left\vert
1\right\rangle \left\vert 0\right\rangle +C_{001}\left( t\right) \left\vert
0\right\rangle \left\vert 0\right\rangle \left\vert 1\right\rangle .
\label{state vector}
\end{eqnarray}

Note that we are not restricting the basis in any way and only considering
the initial conditions leading to single photon and phonon states to
simplify algebra: see, for example, the Appendix in \cite{tokman2021-3}  where arbitrary
multiphoton states are considered in the same way, leading of course to more
cumbersome expressions. Another reason to consider such initial conditions
is that single-photon states are used in most applications, whereas
generating multiphoton number states remains a major challenge.

To determine the anti-Hermitian part $\hat{H}^{\left( ah\right) }$ of the
Hamiltonian which describes relaxation and the correlator $\overline{%
\left\vert \mathfrak{R}\left( t^{\prime }\right) \right\rangle \left\langle
\mathfrak{R}\left( t^{\prime \prime }\right) \right\vert }$ describing
fluctuations, we will use the expression for the total Lindbladian of the
system including a 2-level ``atom'', photons, and phonons \cite{scully1997}. For
simplicity we will assume zero temperature of dissipative reservoirs, which
is satisfied at $T\ll \hbar \omega $. The finite-temperature expressions are
given in \cite{tokman2021,chen2021} . Then the Lindbladian is
\begin{equation}
L\left( \hat{\rho}\right) =L_{e}\left( \hat{\rho}\right) +L_{em}\left( \hat{%
\rho}\right) +L_{p}\left( \hat{\rho}\right)  \label{L phro}
\end{equation}%
\begin{equation}
L_{e}\left( \hat{\rho}\right) =-\frac{\gamma }{2}\left( \hat{\sigma}%
^{\dagger }\hat{\sigma}\hat{\rho}+\hat{\rho}\hat{\sigma}^{\dagger }\hat{%
\sigma}-2\hat{\sigma}\hat{\rho}\hat{\sigma}^{\dagger }\right)
\label{La phro}
\end{equation}%
\begin{equation}
L_{em}\left( \hat{\rho}\right) =-\frac{\mu _{\omega }}{2}\left( \hat{c}%
^{\dagger }\hat{c}\hat{\rho}+\hat{\rho}\hat{c}\hat{c}^{\dagger }-2\hat{c}%
\hat{\rho}\hat{c}^{\dagger }\right)  \label{Lem phro}
\end{equation}%
\begin{equation}
L_{p}\left( \hat{\rho}\right) =-\frac{\mu _{\Omega }}{2}\left( \hat{b}%
^{\dagger }\hat{b}\hat{\rho}+\hat{\rho}\hat{b}\hat{b}^{\dagger }-2\hat{b}%
\hat{\rho}\hat{b}^{\dagger }\right)  \label{Lp phro}
\end{equation}%
where $\gamma $, $\mu _{\omega }$ and $\mu _{\Omega }$ are relaxation rates
of corresponding subsystems.

Then the stochastic equation for the state vector takes the form
\begin{equation}
\left(
\begin{array}{ccc}
\frac{d}{dt} & 0 & 0 \\
0 & \frac{d}{dt}+i\omega _{010}+\gamma _{010} & 0 \\
0 & 0 & \frac{d}{dt}+i\omega _{100}+\gamma _{100}%
\end{array}%
\right) \times \left(
\begin{array}{c}
C_{000} \\
C_{010} \\
C_{100}%
\end{array}%
\right) =-\frac{i}{\hbar }\left(
\begin{array}{c}
\mathfrak{R}_{000} \\
\mathfrak{R}_{010} \\
\mathfrak{R}_{100}%
\end{array}%
\right) ,  \label{stochastic equation for the state vector 1}
\end{equation}

\begin{equation}
\left(
\begin{array}{cc}
\frac{d}{dt}+i\omega _{110}+\gamma _{110} & i\Omega _{R}^{\left( 3\right)
\ast } \\
i\Omega _{R}^{\left( 3\right) } & \frac{d}{dt}+i\omega _{001}+\gamma _{001}%
\end{array}%
\right) \left(
\begin{array}{c}
C_{110} \\
C_{001}%
\end{array}%
\right) =-\frac{i}{\hbar }\left(
\begin{array}{c}
\mathfrak{R}_{110} \\
\mathfrak{R}_{001}%
\end{array}%
\right) ,  \label{stochastic equation for the state vector 1'}
\end{equation}%
where%
\begin{equation*}
\mathfrak{R}_{\alpha ni}=\left\langle \alpha ni\right. \left\vert \mathfrak{R%
}\right\rangle ;
\end{equation*}%
\begin{equation*} 
\omega _{010}=\omega ,\ \omega _{100}=\Omega ,\ \omega _{110}=\omega +\Omega, \ \omega_{001} = \omega_e; 
\end{equation*}%
\begin{equation*}
\gamma _{010}=\frac{1}{2}\mu _{\omega },\ \gamma _{100}=\frac{1}{2}\mu
_{\Omega },\ \gamma _{110}=\frac{1}{2}\left( \mu _{\omega }+\mu _{\Omega
}\right), \ \gamma_{001} = \frac{1}{2} \gamma .
\end{equation*}%
The correlators of the noise sources in Eqs.~(\ref{stochastic equation for
the state vector 1}) and (\ref{stochastic equation for the state vector 1'})
are
\begin{equation}
\overline{\mathfrak{R}_{\alpha ni}^{\ast }\left( t^{\prime }\right)
\mathfrak{R}_{\beta mj}\left( t^{\prime \prime }\right) }=\hbar
^{2}D_{\alpha ni, \beta mj}(t') \delta \left( t^{\prime }-t^{\prime \prime }\right)
,  \label{correlators of the noise sources}
\end{equation}%
where the quantities $D_{\alpha ni,\beta mj}$ are determined using Eqs.~(\ref%
{correlation properties}) and (\ref{L phro})-(\ref{Lp phro}). For the
diagonal elements of the correlators we obtain
\begin{eqnarray}
D_{110,110} &=&D_{001,001}=0  \notag \\
D_{100,100} &=&\mu _{\omega }\overline{\left\vert C_{110}\right\vert ^{2}}
\label{diagonal elements} \\
D_{010,010} &=&\mu _{\Omega }\overline{\left\vert C_{110}\right\vert ^{2}}
\notag \\
D_{000,000} &=&\gamma \overline{\left\vert C_{001}\right\vert ^{2}}+\mu
_{\omega }\overline{\left\vert C_{010}\right\vert ^{2}}+\mu _{\Omega }%
\overline{\left\vert C_{100}\right\vert ^{2}}.  \notag
\end{eqnarray}%
The off-diagonal elements are given by
\begin{eqnarray}
D_{\alpha ni,\beta mj} &=&D_{\beta mj,\alpha ni}^{\ast }  \notag \\
D_{110,\alpha ni} &=&D_{001,\alpha ni}=D_{100,010}=0
\label{off-diagonal elements} \\
D_{000,100} &=&\mu _{\omega }\overline{C_{010}^{\ast }C_{110}}  \notag \\
D_{000,010} &=&\mu _{\Omega }\overline{C_{100}^{\ast }C_{110}}.  \notag
\end{eqnarray}

The dependence of quantities $D_{\alpha ni,\beta mj}$ in the right-hand side
of Eq.~(\ref{correlators of the noise sources}) on time $t^{\prime }$ is due
to the time dependence of amplitudes $C_{\alpha ni}$ which enter Eqs.~(\ref%
{diagonal elements}) and (\ref{off-diagonal elements}).

The derivation of the stochastic equation for the state vector including
pure dephasing processes and the finite temperature of the reservoirs has
been discussed in \cite{tokman2021,chen2021}.

Eqs.~(\ref{stochastic equation for the state vector 1'}) describe the
dynamic generation of an entangled state of the type $\left\vert
MIX\right\rangle =A\left( t\right) \left\vert 0\right\rangle \left\vert
0\right\rangle \left\vert 1\right\rangle +B\left( t\right) \left\vert
1\right\rangle \left\vert 1\right\rangle \left\vert 0\right\rangle $ whereas
Eqs.~(\ref{stochastic equation for the state vector 1}) describe the
relaxation dynamics leading to relaxation of populations to states with
lower energies. The quantities $D_{010,010}$ and $D_{100,100}$ in Eqs.~(\ref%
{diagonal elements}) are associated with processes of the relaxation to
states $\left\vert 0\right\rangle \left\vert 1\right\rangle \left\vert
0\right\rangle $ and $\left\vert 1\right\rangle \left\vert 0\right\rangle
\left\vert 0\right\rangle $ from the entangled state $\left\vert
MIX\right\rangle $ . The structure of the expression for $D_{000,000}$
corresponds to the relaxation of the system from the entangled state to the
ground state via both ``direct'' pathway $%
\left\vert 0\right\rangle \left\vert 0\right\rangle \left\vert
1\right\rangle \rightarrow \left\vert 0\right\rangle \left\vert
0\right\rangle \left\vert 0\right\rangle $ and multistep pathways $%
\left\vert 1\right\rangle \left\vert 1\right\rangle \left\vert
0\right\rangle \rightarrow \left\vert 0\right\rangle \left\vert
1\right\rangle \left\vert 0\right\rangle \rightarrow \left\vert
0\right\rangle \left\vert 0\right\rangle \left\vert 0\right\rangle $ and $%
\left\vert 1\right\rangle \left\vert 1\right\rangle \left\vert
0\right\rangle \rightarrow \left\vert 1\right\rangle \left\vert
0\right\rangle \left\vert 0\right\rangle \rightarrow \left\vert
0\right\rangle \left\vert 0\right\rangle \left\vert 0\right\rangle $, as illustrated in Figs.~3 and 5 below. 

\subsection{Expressions for noise sources in the stochastic equation}

In addition to the expressions for noise correlators, it is 
convenient to know more detailed expressions for the random functions describing the
noise sources $\mathfrak{R}_{\alpha ni}(t)$.

The effect of the reservoir on the dynamic system is characterized by the
matrix elements of the operator which determines coupling to the reservoir.
For a weak coupling these matrix elements are linear with respect to the
matrix elements of the operators describing the dynamical system. Therefore, the functions $\mathfrak{R}_{\alpha ni}(t)$ should depend linearly on the components of the state vector of the system. 

Here we again consider the low-temperature case when the relaxation
processes can bring the populations only down, not up. When the reservoirs
for each subsystem are statistically independent, one can try the following ansatz, 
\begin{equation}
\mathfrak{R}_{\alpha ni}\left( t\right) =\hbar \sqrt{\gamma }C_{\alpha
n\left( i+1\right) }\left( t\right) f_{e}\left( t\right) +\hbar \sqrt{\mu
_{\omega }}C_{\alpha \left( n+1\right) i}\left( t\right) f_{em}\left(
t\right) +\hbar \sqrt{\mu _{\Omega }}C_{\left( \alpha +1\right) ni}\left(
t\right) f_{p}\left( t\right) ,  \label{noise sources in low T}
\end{equation}%
where $\overline{f_{e,em,p}}=0.$ Here $f_{e,em,p}\left( t\right) $ are
random functions that are determined by the statistics of noise in
unperturbed electron, photon, and phonon reservoirs. The functions $f_{e,em,p}$ should not depend on the
set of variables $C_{\alpha ni}$ within the above approximations.

The linear dependence of the noise term on the state vector was also assumed
in SLE \cite{kostin1972,katz2016}, in which the noise terms had the form
\begin{equation}
\left\vert \mathfrak{R}\right\rangle =\hat{U}\left( \mathbf{r},t\right)
\left\vert \Psi \right\rangle ,  \label{noise terms}
\end{equation}
where $\hat{U}\left( \mathbf{r},t\right) $ is the fluctuating component of
the potential.

To ensure that Eq.~(\ref{noise sources in low T}) leads to the correlators Eqs.~(\ref%
{correlators of the noise sources})-(\ref{off-diagonal elements}) that are
consistent with the Lindblad master equation, one needs first to define the
correlators of the random functions in Eq.~(\ref{noise sources in low T}) in the
following way:
\begin{equation}
\overline{f_{\kappa }^{\ast }\left( t^{\prime }\right) f_{\lambda }\left(
t^{\prime \prime }\right) }=\delta _{\kappa \lambda }\delta \left( t^{\prime
}-t^{\prime \prime }\right) ,  \label{f con}
\end{equation}%
where $\kappa ,\lambda =e,em,p$, i.e. the fluctuations in different
reservoirs are independent and Markovian. Second, one has to assume that the
correlations are factorized when calculating the averages
\begin{equation}
\overline{C_{\alpha ni}^{\ast }\left( t^{\prime }\right) C_{\beta mj}\left(
t^{\prime \prime }\right) f_{\kappa }^{\ast }\left( t^{\prime }\right)
f_{\lambda }\left( t^{\prime \prime }\right) }=\overline{C_{\alpha ni}^{\ast
}\left( t^{\prime }\right) C_{\beta mj}\left( t^{\prime \prime }\right) }%
\times \overline{f_{\kappa }^{\ast }\left( t^{\prime }\right) f_{\lambda
}\left( t^{\prime \prime }\right) }.  \label{factorization}
\end{equation}

Eq.~(\ref{f con}) looks obvious, whereas the factorization in Eq.~(\ref%
{factorization}) is valid only in linear approximation with respect to
relaxation constants $\gamma $ and $\mu _{\omega ,\Omega }$. However, the
Lindbladian of the form given in Eqs.~(\ref{state vector})-(\ref{Lem phro})
is itself valid within the same approximation. Therefore, Eqs.~(\ref{noise
sources in low T}), (\ref{f con}), and (\ref{factorization}) lead to all expressions in
Eqs.~(\ref{diagonal elements}),(\ref{off-diagonal elements}). One also has to 
keep in mind that with our choice of our initial conditions the amplitudes $C_{\alpha ni}=0$ for all states with energies
above those in states $\left\vert 1\right\rangle \left\vert 1\right\rangle
\left\vert 0\right\rangle $ and $\left\vert 0\right\rangle \left\vert
0\right\rangle \left\vert 1\right\rangle $.

%%%%%%%%%%%%%%%%%%%%%%%%%%%

\section{Dynamics of entangled fermion-photon-phonon states in a
dissipative system}

Here we write an explicit solution of Eqs~(\ref{stochastic equation for the
state vector 1}) and (\ref{stochastic equation for the state vector 1'}) for
the initial state vector $\left\vert \Psi \left( 0\right) \right\rangle
=\left\vert 0\right\rangle \left\vert 0\right\rangle \left\vert
1\right\rangle $, when $C_{001}\left( 0\right) =1$, $C_{000}\left( 0\right)
=C_{010}\left( 0\right) =C_{100}\left( 0\right) =C_{110}\left( 0\right) =0$.
Assuming exact resonance at $\omega_e=\omega +\Omega $, we obtain
\begin{eqnarray}
C_{000}\left( t\right) &=&-\frac{i}{\hbar }\int_{0}^{t}\mathfrak{R}%
_{000}\left( t^{\prime }\right) dt^{\prime }  \notag \\
C_{010}\left( t\right) &=&-\frac{i}{\hbar }\int_{0}^{t}\mathfrak{R}%
_{010}\left( t^{\prime }\right) e^{-i\omega \left( t-t^{\prime }\right)
-\gamma _{010}\left( t-t^{\prime }\right) }dt^{\prime }  \label{solution1} \\
C_{100}\left( t\right) &=&-\frac{i}{\hbar }\int_{0}^{t}\mathfrak{R}%
_{100}\left( t^{\prime }\right) e^{-i\Omega \left( t-t^{\prime }\right)
-\gamma _{100}\left( t-t^{\prime }\right) }dt^{\prime },  \notag
\end{eqnarray}

\begin{eqnarray}
C_{001}\left( t\right) &=&e^{-i\omega_et}e^{-\frac{\gamma _{110}+\gamma
_{001}}{2}t}\left[ \cos \left( \tilde{\Omega}_{R}t\right) +\frac{\gamma
_{110}-\gamma _{001}}{\tilde{\Omega}_{R}}\sin \left( \tilde{\Omega}%
_{R}t\right) \right] +\delta C_{001}  \label{solution2} \\
C_{110}\left( t\right) &=&e^{-i\omega_et}e^{-\frac{\gamma _{110}+\gamma
_{001}}{2}t}\left( ie^{-i\theta }\right) \sin \left( \tilde{\Omega}%
_{R}t\right) +\delta C_{110}  \label{solution2-2} 
\end{eqnarray}%
where the effective Rabi frequency $\tilde{\Omega}_{R}=\sqrt{\left\vert \Omega _{R}^{\left( 3\right)
}\right\vert ^{2}-\frac{\left( \gamma _{110}-\gamma _{001}\right) ^{2}}{4}}$%
, $\theta =\mathrm{Arg}\left[ \Omega _{R}^{\left( 3\right) }\right] $, and
the terms $\delta C_{100,110}$ are linear with respect to random functions $%
\mathfrak{R}_{001}$ and $\mathfrak{R}_{110}$.

The term proportional to $\frac{\gamma _{110}-\gamma _{001}}{\tilde{\Omega}%
_{R}}$ in the first of Eqs. (\ref{solution2}) can be omitted when
calculating most observables when the dissipation is weak, $\tilde{\Omega}%
_{R}\gg $ $\gamma _{ani}$ (see, e.g., \cite{tokman2021}). However, one has
to keep in mind that this term is needed for Eqs. (\ref{solution2}) to
satisfy an exact integral of motion of Eqs.~(\ref{stochastic equation for
the state vector 1'}) :%
\begin{equation*}
\frac{d}{dt}\left( \overline{\left\vert C_{001}\right\vert ^{2}}+\overline{%
\left\vert C_{110}\right\vert ^{2}}\right) =-2\gamma _{001}\overline{%
\left\vert C_{001}\right\vert ^{2}}-2\gamma _{110}\overline{\left\vert
C_{110}\right\vert ^{2}}.
\end{equation*}%
Note that when averaged over the period $\frac{2\pi }{\tilde{\Omega}_{R}}$
the integral is conserved even without this term.

The solution for the 5-component state vector can be written as
\begin{eqnarray}
&&\left\vert \Psi \right\rangle  \notag \\
&=&e^{-i\omega_et}e^{-\frac{\gamma _{110}+\gamma _{001}}{2}t}\left\{ %
\left[ \cos \left( \tilde{\Omega}_{R}t\right) +\frac{\gamma _{110}-\gamma
_{001}}{2\tilde{\Omega}_{R}}\sin \left( \tilde{\Omega}_{R}t\right) \right]
\left\vert 0\right\rangle \left\vert 0\right\rangle \left\vert
1\right\rangle +ie^{-i\theta }\sin \left( \tilde{\Omega}_{R}t\right)
\left\vert 1\right\rangle \left\vert 1\right\rangle \left\vert
0\right\rangle \right\}  \notag \\
&&+\delta C_{001}\left\vert 0\right\rangle \left\vert 0\right\rangle
\left\vert 1\right\rangle +\delta C_{110}\left\vert 1\right\rangle
\left\vert 1\right\rangle \left\vert 0\right\rangle +C_{000}\left\vert
0\right\rangle \left\vert 0\right\rangle \left\vert 0\right\rangle
+C_{100}\left\vert 1\right\rangle \left\vert 0\right\rangle \left\vert
0\right\rangle +C_{010}\left\vert 0\right\rangle \left\vert 1\right\rangle
\left\vert 0\right\rangle ,
\label{solution for the 5-component state vector}
\end{eqnarray}%
where%
\begin{equation}
\overline{\delta C_{001}}=\overline{\delta C_{110}}=\overline{C_{000}}=%
\overline{C_{100}}=\overline{C_{010}}=0,  \label{avc}
\end{equation}%
\begin{equation}
\overline{\left\vert \delta C_{001}\right\vert ^{2}}=\overline{\left\vert
\delta C_{110}\right\vert ^{2}}=0.  \label{avc2}
\end{equation}

It follows from Eq. (\ref{solution for the 5-component state vector}) that
in the entangled state $\left\vert MIX\right\rangle =A\left( t\right)
\left\vert 0\right\rangle \left\vert 0\right\rangle \left\vert
1\right\rangle +B\left( t\right) \left\vert 1\right\rangle \left\vert
1\right\rangle \left\vert 0\right\rangle $ the amplitudes $A\left( t\right) $
and $B\left( t\right) $ oscillate at the effective Rabi frequency and decay with the
decay rate
\begin{equation*}
\gamma _{_{MIX}}=\frac{\gamma _{110}+\gamma _{001}}{2}=\frac{1}{4}\left( \mu
_{\omega }+\mu _{\Omega }+\gamma \right) .
\end{equation*}

The occupation probabilities $|C_{001}|^2$ and $|C_{110}|^2$ are plotted in Fig.~2 as a function of normalized time $\tilde{\Omega}_{R}t$, along with the real parts of their eigenfrequencies obtained from Eqs.~(\ref{stochastic equation for
the state vector 1'}) as a function of detuning from the parametric resonance $\omega + \Omega - \omega_e$. Although the plots look like standard anticrossing behavior and decaying Rabi oscillations, one should keep in mind that (1) the anticrossing occurs not at the standard exciton-photon or phonon-photon resonance but at the nonlinear parametric resonance, which is controlled by the nonlinear coupling strength $\Omega_{R}^{(3)}$ and entangles three degrees of freedom; (2) the relaxation rates of each individual subsystem enter the analytic expressions plotted in Fig.~2 in a nontrivial way, as described above. This dependence will become even more nontrivial at high temperature of the reservoir. The presented solution provides the way to retrieve the analytic  dependence of any observable on the relaxation and coupling parameters and determine correctly the criterion for observing the strong parametric coupling in frequency or time domain.  Two obvious examples for such observables are photon and phonon emission spectra that are derived and plotted in Sec.~VI, see Figs.~4 and 6. 

%%%%%%%%%%%%%%%%%%%%%%%%%%%%%%%%%%%%%

\begin{figure}[htb]
\centering
\begin{subfigure}[b]{0.5\textwidth}
\includegraphics[width=1\linewidth]{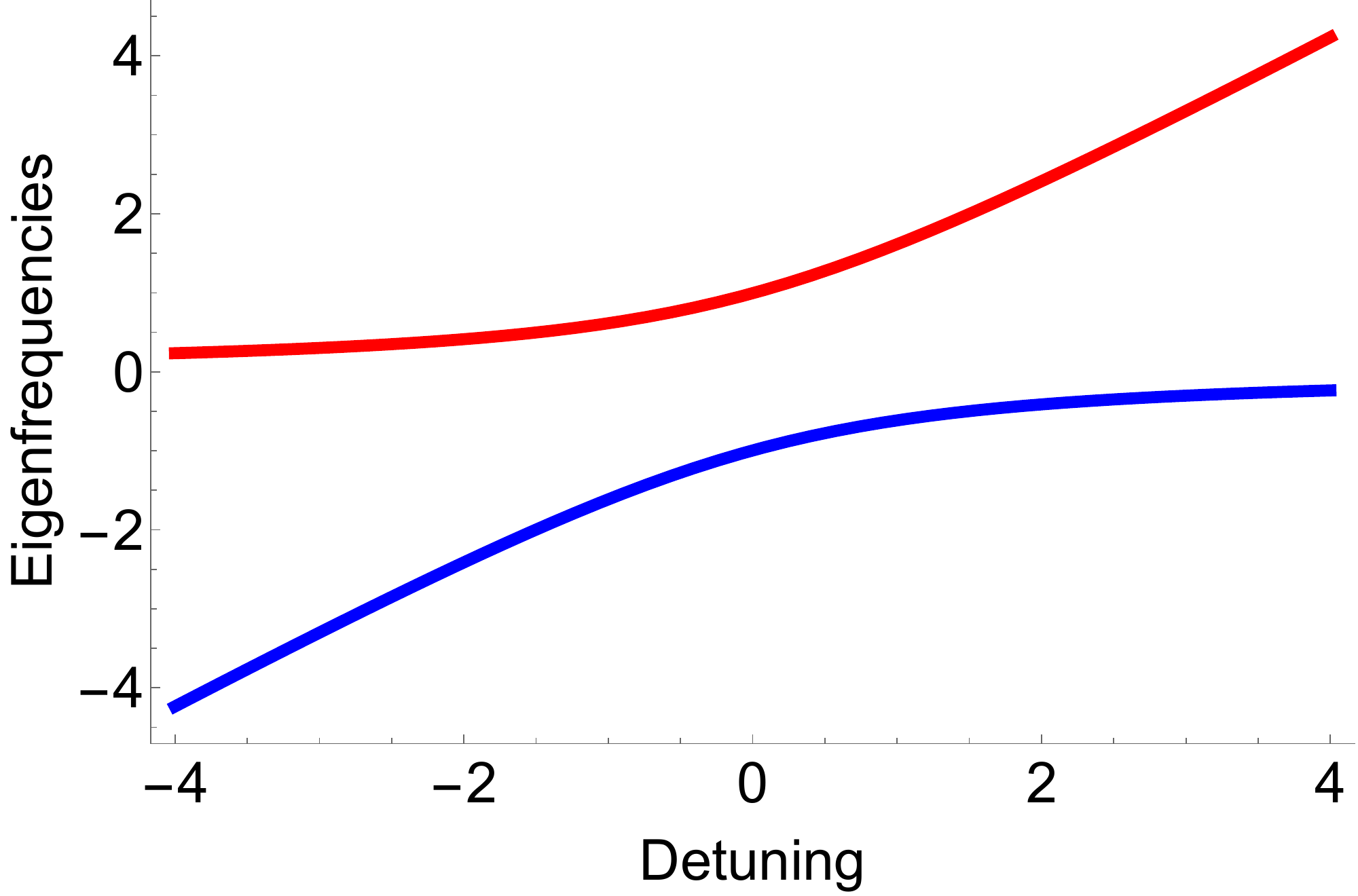}
\caption{}
\label{fig2a}
\end{subfigure}

\begin{subfigure}[b]{0.5\textwidth}
\includegraphics[width=1\linewidth]{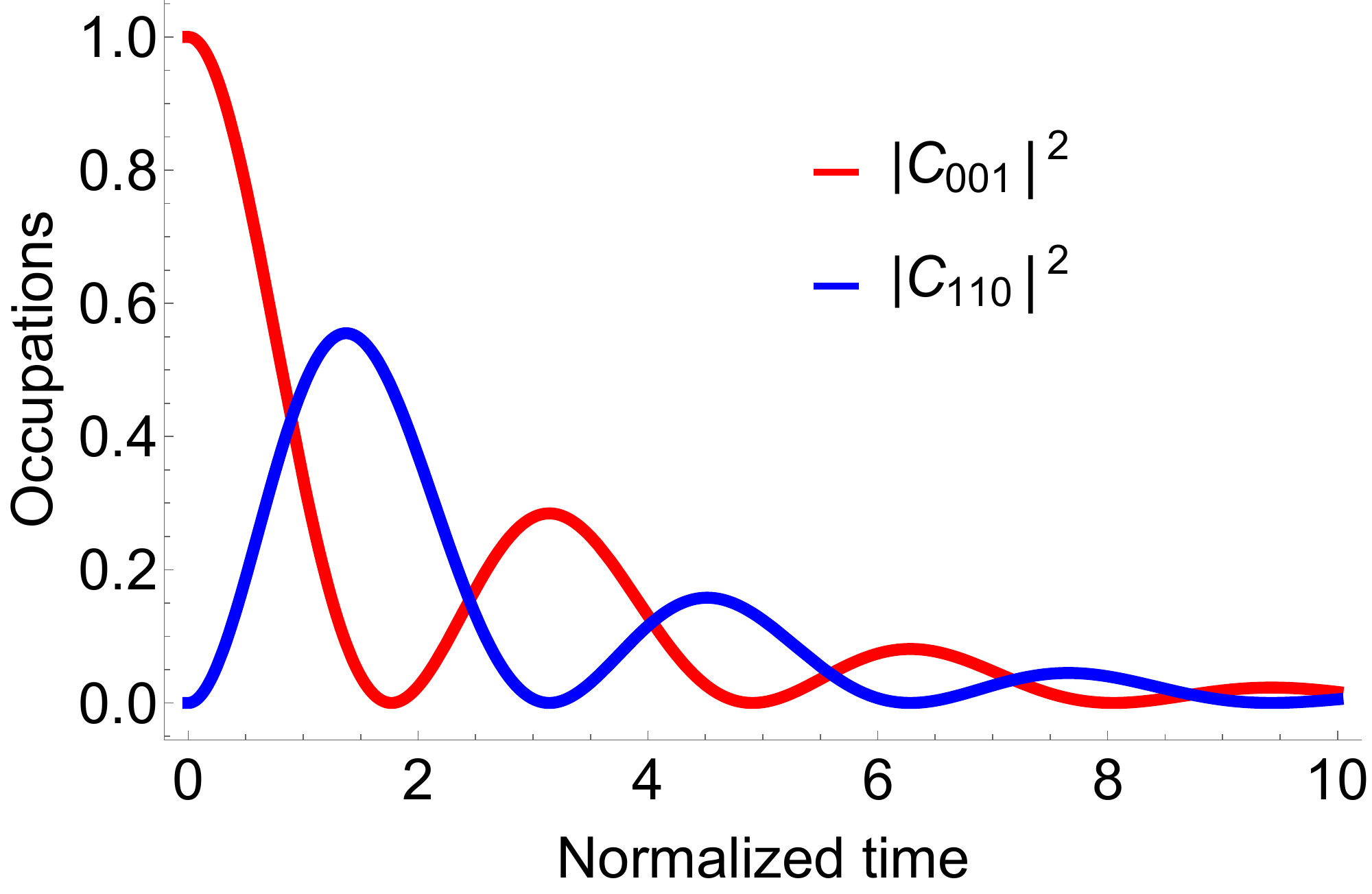}
\caption{}
\label{fig2b}
\end{subfigure}
\caption{  (a) Real parts of eigenstate frequencies of Eqs.~(\ref{stochastic equation for
the state vector 1'}), shifted by the electron transition frequency $\omega_e$ and normalized by $|\Omega _{R}^{(3)}|$, as a function of detuning from the parametric resonance $\omega + \Omega - \omega_e$ normalized by $|\Omega _{R}^{(3)}| $. The relaxation rates are $\mu_{\omega} = \mu_{\Omega}  = 0.3 |\Omega _{R}^{(3)}| $ and $\gamma = 0.2 |\Omega _{R}^{(3)}| $. 
(b)  Occupation probabilities $|C_{001}|^2$ and $|C_{110}|^2$ from Eqs.~(\ref{solution2}) and (\ref{solution2-2}) as a function of normalized time $\tilde{\Omega}_{R} t$ for the same relaxation rates.   }
\end{figure}

%%%%%%%%%%%%%%%%%%%%%%%%%

We also give the expressions for the occupation probabilities $\overline{\left\vert
C_{100}\right\vert ^{2}}$, $\overline{\left\vert C_{010}\right\vert ^{2}}$
and $\overline{\left\vert C_{000}\right\vert ^{2}}$ that are valid under the
condition $\tilde{\Omega}_{R}\gg $ $\gamma _{ani}$:%
\begin{equation}
\overline{\left\vert C_{100}\right\vert ^{2}}\approx \frac{\mu _{\omega }}{%
\mu _{\Omega }-\mu _{\omega }-\gamma }\left( e^{-\frac{\mu _{\Omega }+\mu
_{\omega }+\gamma }{2}t}-e^{-\mu _{\Omega }t}\right)  \label{c100 2}
\end{equation}%
\begin{equation}
\overline{\left\vert C_{010}\right\vert ^{2}}\approx \frac{\mu _{\Omega }}{%
\mu _{\omega }-\mu _{\Omega }-\gamma }\left( e^{-\frac{\mu _{\Omega }+\mu
_{\omega }+\gamma }{2}t}-e^{-\mu _{\omega }t}\right)  \label{c010 2}
\end{equation}%
\begin{equation}
\overline{\left\vert C_{000}\right\vert ^{2}}\approx \frac{\gamma \left(
\gamma -\mu _{\Omega }-\mu _{\omega }\right) }{\gamma ^{2}-\left( \mu
_{\Omega }-\mu _{\omega }\right) ^{2}}\left( 1-e^{-\frac{\mu _{\Omega }+\mu
_{\omega }+\gamma }{2}t}\right) -\left( \mu _{\Omega }\frac{1-e^{-\mu
_{\omega }t}}{\mu _{\omega }-\mu _{\Omega }-\gamma }+\mu _{\omega }\frac{%
1-e^{-\mu _{\Omega }t}}{\mu _{\Omega }-\mu _{\omega }-\gamma }\right)
\label{c000 2}
\end{equation}%
It is easy to see that Eqs.~(\ref{c100 2})-(\ref{c000 2}) don't have the
divergence when $\left[ \pm \left( \mu _{\omega }-\mu _{\Omega }\right)
-\gamma \right] \longrightarrow 0$; for example,
\begin{equation*}
\lim_{\left( \mu _{\Omega }-\mu _{\omega }-\gamma \right) \longrightarrow 0}
\left[ \frac{e^{-\frac{\mu _{\Omega }+\mu _{\omega }+\gamma }{2}t}-e^{-\mu
_{\Omega }t}}{\mu _{\Omega }-\mu _{\omega }-\gamma }\right] =\frac{1}{2}%
te^{-\mu _{\Omega }t}
\end{equation*}%
and so on. In the limit $t\longrightarrow \infty $ Eqs.~(\ref{c100 2})-(\ref%
{c000 2}) lead to $\overline{\left\vert C_{100}\right\vert ^{2}}=\overline{%
\left\vert C_{010}\right\vert ^{2}}=0$, $\overline{\left\vert
C_{000}\right\vert ^{2}}=1$.

The detailed derivation of Eqs.~(\ref{c100 2})-(\ref{c000 2}) is given in
Appendix A. It is also shown there that in our case the quantities $\delta
C_{100}$ and $\delta C_{110}$ do not contribute to the observables and
therefore can be omitted.

In \cite{tokman2021} we used a simplified model to analyze the
fermion-photon-phonon entanglement, in which all relaxation pathways to the
ground state $\left\vert 0\right\rangle \left\vert 0\right\rangle \left\vert
0\right\rangle $ are assumed to be ``
direct'', which corresponds to taking all correlators equal
to zero except $D_{000,000}$. This approach is essentially the
Weisskopf-Wigner method, modified in order to conserve the norm of the state
vector. It gives a correct result for the decay rate $\gamma _{MIX}$ of
the entangled state. At the same time, including multistep decay pathways is
of principal importance when calculating and interpreting the emission spectra, as we will
see in the next section.

%%%%%%%%%%%%%%%%%%%%%%%%%%%%%%%%%%%%

\section{Emission spectra of photons and phonons from the
parametric decay of the electron excitation}

\subsection{ Derivation of the emission spectra from the
solution of the stochastic equation for the state vector: a general scheme}

Consider for definiteness the EM radiation out of a cavity. Its
power spectrum received by the detector is given by \cite%
{scully1997,madsen2013}%
\begin{equation*}
P\left( \nu \right) =A\cdot S\left( \nu \right) ,
\end{equation*}

where%
\begin{equation}
S\left( \nu \right) =\frac{1}{\pi }\mathrm{Re}\int_{0}^{\infty }d\tau
e^{i\nu \tau }\int_{0}^{\infty }dtK\left( t,\tau \right) ,  \label{S(v)}
\end{equation}%
$K=\left\langle \hat{c}^{\dagger }\left( t\right) \hat{c}\left( t+\tau
\right) \right\rangle $, $\left\langle \cdots \right\rangle $ is a
quantum-mechanical averaging. The coefficient A includes the $Q$-factor of a
cavity, spatial structure of the outgoing field, and the position and
properties of the detector.

To calculate the power spectrum one needs to know the solution of the
Heisenberg-Langevin equations for the field operators $\hat{c}\left(
t\right) $ and $\hat{c}^{\dagger }\left( t\right) $, then calculate the
correlator, and average it over the statistics of Langevin noise: $%
K\Rightarrow $ $\overline{\left\langle \hat{c}^{\dagger }\left( t\right)
\hat{c}\left( t+\tau \right) \right\rangle }$. However, as we already
discussed, the Heisenberg-Langevin equations become nonlinear in the
strong-coupling regime. Therefore, it may be more convenient to obtain the
spectra from the solution of the stochastic equation Eq.(33) for the state
vector. The general procedure is as follows.

First, we need to transform the correlator%
\begin{equation}
K\left( t,\tau \right) =\left\langle \hat{c}^{\dagger }\left( t\right) \hat{c%
}\left( t+\tau \right) \right\rangle =\left\langle \Psi \left( 0\right)
\right\vert \hat{c}^{\dagger }\left( t\right) \hat{c}\left( t+\tau \right)
\left\vert \Psi \left( 0\right) \right\rangle   \label{k(t,tao)}
\end{equation}%
to the Schr\"{o}dinger picture without taking into account dissipation and
fluctuations. If $\hat{U}\left( t\right) $ is the unitary operator of
evolution of the system, one can write
\begin{equation}
K=\left\langle \Psi \left( 0\right) \right\vert \hat{U}^{\dagger }\left(
t\right) \hat{c}^{\dagger }\, \hat{U}\left( t\right) \hat{U}^{\dagger }\left(
t+\tau \right) \hat{c}\, \hat{U}\left( t+\tau \right) \left\vert \Psi \left(
0\right) \right\rangle =\left\langle \hat{c}\, \Psi \left( t\right) \right\vert
\hat{U}\left( t\right) \hat{U}^{\dagger }\left( t+\tau \right) \left\vert
\hat{c}\, \Psi \left( t+\tau \right) \right\rangle ,  \label{k}
\end{equation}%
where $\hat{c}$ is the Schr\"{o}dinger's (constant) operator which we will treat as an
initial condition for the Heisenberg operator $\hat{c}\left( t\right) $ at $%
t=0$. We will use the notation $\hat{U}\left( t\right) \equiv \hat{U}%
_{t_{0}}\left( t^{\prime }\right) $, where we indicate explicitly the
initial moment of time $t_{0}$ and the duration of evolution $t^{\prime
}=t-t_{0}$. This will lead to the following replacements in Eq.~(\ref{k}): $%
\hat{U}\left( t\right) \Longrightarrow \hat{U}_{0}\left( t\right) $, $\hat{U}%
\left( t+\tau \right) \Longrightarrow \hat{U}_{0}\left( t+\tau \right) $.
Furthermore, we obviously have
\begin{equation}
\hat{U}_{0}\left( t+\tau \right) =\hat{U}_{t}\left( \tau \right) \hat{U}%
_{0}\left( t\right)  \label{U(t+tau)}
\end{equation}%
which gives
\begin{equation*}
K=\left\langle \hat{c}\,\Psi \left( t\right) \right\vert \hat{U}_{0}\left(
t\right) \left( \hat{U}_{t}\left( \tau \right) \hat{U}_{0}\left( t\right)
\right) ^{\dagger }\left\vert \hat{c}\, \Psi \left( t+\tau \right)
\right\rangle =\left\langle \hat{c}\, \Psi \left( t\right) \right\vert \hat{U}%
_{0}\left( t\right) \hat{U}_{0}^{\dagger }\left( t\right) \hat{U}%
_{t}^{\dagger }\left( \tau \right) \left\vert \hat{c}\, \Psi \left( t+\tau
\right) \right\rangle .
\end{equation*}%
Taking into account $\hat{U}_{0}\left( t^{\prime }\right) \hat{U}%
_{0}^{\dagger }\left( t^{\prime }\right) =1$, we obtain
\begin{equation}
K=\left\langle \hat{U}_{t}\left( \tau \right) \hat{c}\,  \Psi \left( t\right)
\right. \left\vert \hat{c}\, \Psi \left( t+\tau \right) \right\rangle .
\label{cor}
\end{equation}%
Second, introducing the notations
\begin{equation}
\Psi _{\hat{C}}\left( t\right) =\hat{c}\, \Psi \left( t\right) ,\ \ \Phi \left(
t,\tau \right) =\hat{U}_{t}\left( \tau \right) \Psi _{\hat{C}}\left(
t\right) ,  \label{notations}
\end{equation}%
we arrive at
\begin{equation}
K\left( t,\tau \right) =\left\langle \Phi \left( t,\tau \right) \right.
\left\vert \Psi _{\hat{C}}\left( t+\tau \right) \right\rangle .
\label{K(t,tao)}
\end{equation}

Therefore, in order to calculate the correlator $K$ through the solution of
the equation for the state vector, one has to perform the following steps:

\textbf{a)} Find vector $\left\vert \Psi _{\hat{C}}\left( t+\tau \right)
\right\rangle =$ $\hat{c}\Psi \left( t+\tau \right) $, where $\Psi \left(
t+\tau \right) $ is the solution for the state vector $\left\vert \Psi
\right\rangle $ at the time interval $\left[ 0,t+\tau \right] $ with initial
condition $\left\vert \Psi \left( 0\right) \right\rangle $.

\textbf{b)} Find vector $\left\vert \Phi \left( t,\tau \right) \right\rangle
$. To do that, one has to solve for the state vector $\left\vert \Psi
\right\rangle $ at the time interval $\left[ t,t+\tau \right] $ with initial
condition $\left\vert \Psi _{\hat{C}}\left( t\right) \right\rangle $. The
vector $\left\vert \Psi _{\hat{C}}\left( t\right) \right\rangle $ is the
same as in part a), but instead of the time interval $\left[ 0,t+\tau \right]
$ one has to take the time interval $\left[ 0,t\right] $.

To check Eq.~(\ref{K(t,tao)}) for consistency, we note that in the absence
of dissipation one can go back from this equation to the standard expression
which follows directly from the initial equation (\ref{k(t,tao)}):
\begin{equation*}
K\left( t,\tau \right) =\left\langle \Psi \left( 0\right) \right\vert e^{i%
\frac{\hat{H}}{\hbar }t}\hat{c}^{\dagger }e^{i\frac{\hat{H}}{\hbar }\tau }%
\hat{c}e^{-i\frac{\hat{H}}{\hbar }\left( t+\tau \right) }\left\vert \Psi
\left( 0\right) \right\rangle .
\end{equation*}

If the dynamic system is open, then a complete closed system
``the dynamic system + reservoir'' has its
own unitary operator of evolution $\hat{U}_{t_{0}}\left( t^{\prime }\right) $%
. Therefore, Eq.~(\ref{K(t,tao)}) should be valid for a complete system as
well which includes the reservoir variables. Now we apply the Langevin
method which assumes that the averaging over the statistics of noise sources
entering a stochastic equation (in this case Eq.~(\ref{stochastic eq for psi}%
)) is equivalent to averaging over the reservoir variables. Therefore, we
can solve Eq.~(\ref{stochastic eq for psi}) and, following the above steps,
find the functions $\Psi _{\hat{C}}\left( t\right) $, $\Psi _{\hat{C}}\left(
t+\tau \right) $ and $\Phi\left( t,\tau \right) $ which
are now dependent on the noise sources. Then we substitute the latter two
functions into Eq.~(\ref{K(t,tao)}) and perform averaging over the noise
statistics. As a result, we obtain
\begin{equation}
K\left( t,\tau \right) =\overline{\left\langle \ \Phi \left( t,\tau \right)
\right. \left\vert \Psi _{\hat{C}}\left( t+\tau \right) \right\rangle }.
\label{k bar}
\end{equation}

%%%%%%%%%%%%%%%%%%%%%%%%%%%%%%%%%%

\subsection{Photon emission spectra for the parametric decay of an excited
electron }

Here we apply the general recipe of calculating $K\left( t,\tau \right) $
formulated in the previous section to a particular example of the parametric
decay of an initially excited fermionic two-level system under strong
coupling to a parametric electron-photon-phonon resonance.

a) Use the expressions (\ref{solution1})-(\ref{solution for the
5-component state vector}) to find the vector $\left\vert \Psi \left(
t\right) \right\rangle $.

b) Determine vectors $\left\vert \Psi _{\hat{C}}\left( t\right)
\right\rangle $ and $\left\vert \Psi _{\hat{C}}\left( t+\tau \right)
\right\rangle $, resulting in
\begin{equation}
\left\vert \Psi _{\hat{C}}\left( t\right) \right\rangle =\hat{c}\left\vert
\Psi \left( t\right) \right\rangle =C_{110}\left( t\right) \left\vert
1\right\rangle \left\vert 0\right\rangle \left\vert 0\right\rangle
+C_{010}\left( t\right) \left\vert 0\right\rangle \left\vert 0\right\rangle
\left\vert 0\right\rangle,  \label{phsi t}
\end{equation}%
\begin{equation}
\left\vert \Psi _{\hat{C}}\left( t+\tau \right) \right\rangle =\hat{c}%
\left\vert \Psi \left( t+\tau \right) \right\rangle =C_{110}\left( t+\tau
\right) \left\vert 1\right\rangle \left\vert 0\right\rangle \left\vert
0\right\rangle +C_{010}\left( t+\tau \right) \left\vert 0\right\rangle
\left\vert 0\right\rangle \left\vert 0\right\rangle.   \label{phsi t+tau}
\end{equation}

c) To determine the vector $\left\vert \Phi \left( t,\tau \right)
\right\rangle $ we will use the solution of Eqs.~(\ref{stochastic equation
for the state vector 1}),(\ref{stochastic equation for the state vector 1'})
at the time interval $\left[ t,t+\tau \right] $ where the initial condition $%
\left\vert \Psi _{\hat{C}}\left( t\right) \right\rangle $ is given by Eq.~(%
\ref{phsi t}). In our case Eq.~(\ref{phsi t}) determines the initial value
of the state vector; its subsequent evolution is determined by a simple Eq.~(%
\ref{stochastic equation for the state vector 1}) for the amplitudes of
states $\left\vert 1\right\rangle \left\vert 0\right\rangle \left\vert
0\right\rangle $ and $\left\vert 0\right\rangle \left\vert 0\right\rangle
\left\vert 0\right\rangle $. As a result, vector $\left\vert \Phi \left(
t,\tau \right) \right\rangle $ is given by
\begin{eqnarray}
\left\vert \Phi \left( t,\tau \right) \right\rangle &=&C_{100}^{\left( \Phi
\right) }\left( t,\tau \right) \left\vert 1\right\rangle \left\vert
0\right\rangle \left\vert 0\right\rangle +C_{000}^{\left( \Phi \right)
}\left( t,\tau \right) \left\vert 0\right\rangle \left\vert 0\right\rangle
\left\vert 0\right\rangle  \notag \\
&=&\left( e^{-i\omega \tau -\gamma _{100}\tau }C_{110}\left( t\right) -\frac{%
i}{\hbar }\int_{t}^{t+\tau }\mathfrak{R}_{100}^{\left( \Phi \right) }\left(
t,t^{\prime }\right) e^{-i\omega \left( \tau +t-t^{\prime }\right) -\gamma
_{100}\left( \tau +t-t^{\prime }\right) }dt^{\prime }\right) \left\vert
1\right\rangle \left\vert 0\right\rangle \left\vert 0\right\rangle  \notag \\
&&+\left( C_{010}\left( t\right) -\frac{i}{\hbar }\int_{t}^{t+\tau }%
\mathfrak{R}_{000}^{\left( \Phi \right) }\left( t,t^{\prime }\right)
dt^{\prime }\right) \left\vert 0\right\rangle \left\vert 0\right\rangle
\left\vert 0\right\rangle ,  \label{phi (t tau)}
\end{eqnarray}%
where functions $C_{110}\left( t\right) $ and $C_{010}\left( t\right) $ are
determined by Eqs. (\ref{solution1}), (\ref{solution2}).

The superscript $\left( \Phi \right) $ in the terms $\mathfrak{R}_{\alpha
ni} $ in Eq.~(\ref{phi (t tau)}) means that the correlators of these noise
terms correspond to the state vector $\left\vert \Phi \right\rangle $. The
dependence on the initial time moment $t$ of the evolution in $\mathfrak{R}%
_{\alpha ni}^{\left( \Phi \right) }\left( t,t^{\prime }\right) $ takes into
account that the correlators of these random functions may depend on the
value of $t$ as a parameter, because complex amplitudes $C_{\alpha
ni}^{\left( \Phi \right) }\left( t,\tau \right) \equiv C_{\alpha ni}^{\left(
\Phi \right) }\left( t,t^{\prime }-t\right) $ depend on this parameter.

Next, we substitute the expressions for $C_{110}\left( t\right) $, $%
C_{010}\left( t\right) $, $C_{110}\left( t+\tau \right) $ determined by Eqs.
(\ref{solution1}) and (\ref{solution2}), into Eqs. (\ref{phsi t+tau}) and (%
\ref{phi (t tau)}); after that we substitute Eqs. (\ref{phsi t+tau}) and (%
\ref{phi (t tau)}) into Eq.~(\ref{k bar}). In the resulting expression we
average over the noise statistics taking into account that the noise sources
are delta-correlated. Omitting the terms that become zero after averaging,
we obtain
\begin{eqnarray}
K\left( t,\tau \right) &=&e^{-i\omega \tau -\left( \gamma _{100}+\frac{%
\gamma _{110}+\gamma _{001}}{2}\right) \tau -\left( \gamma _{110}+\gamma
_{001}\right) t}\sin \left( \tilde{\Omega}_{R}t\right) \sin \left[ \tilde{%
\Omega}_{R}\left( t+\tau \right) \right]  \notag \\
&&+e^{-i\omega \tau -\gamma _{010}\tau -2\gamma
_{010}t}\int_{0}^{t}D_{010,010}\left( t^{\prime }\right) e^{2\gamma
_{010}t^{\prime }}dt^{\prime }  \notag \\
&&+e^{-i\omega \left( t+\tau \right) -\gamma _{010}\left( t+\tau \right)
-2\gamma _{010}t}\int_{t}^{t+\tau }\tilde{D}_{000,010}\left( t,t^{\prime
}\right) e^{\left( i\omega +\gamma _{010}\right) t^{\prime }}dt^{\prime }.
\label{K t, tau}
\end{eqnarray}%
Here the quantity $D_{010,010}$ is determined by Eqs.~(\ref{diagonal
elements}):
\begin{equation}
D_{010,010}=2\gamma _{010}\overline{\left\vert C_{110}\left( t^{\prime
}\right) \right\vert ^{2}}=\mu _{\Omega }\overline{\left\vert C_{110}\left(
t^{\prime }\right) \right\vert ^{2}}.  \label{d 010010}
\end{equation}%
The function $\tilde{D}_{000,010}\left( t,t^{\prime }\right) $ corresponds
to the following correlator:%
\begin{equation}
\overline{\mathfrak{R}_{000}^{\left( \Phi \right) \ast }\left( t,t^{\prime
}\right) \mathfrak{R}_{010}\left( t^{\prime \prime }\right) }=\hbar ^{2}%
\tilde{D}_{000,010}\left( t,t^{\prime }\right) \delta \left( t^{\prime
}-t^{\prime \prime }\right).  \label{correlator}
\end{equation}%
To calculate the value of $\tilde{D}_{000,010}\left( t,t^{\prime }\right) $
it is not enough to have expressions (\ref{diagonal elements}) and (\ref%
{off-diagonal elements}) because it is determined by correlations between
the noise terms for \textit{different} state vectors $\left\vert \Phi
\right\rangle $ and $\left\vert \Psi \right\rangle $, which correspond to
the solutions of the equation for the state vector with \textit{different
initial conditions}. We need to use the expression for the noise source
obtained in Sec.~IV.D. From Eqs.~(\ref{noise
sources in low T}), (\ref{f con}), and (\ref{factorization}) we obtain
\begin{equation}
\tilde{D}_{000,010}\left( t,t^{\prime }\right) =\mu _{\Omega }\overline{%
C_{100}^{\left( \Phi \right) \ast }\left( t,t^{\prime }-t\right)
C_{110}\left( t^{\prime }\right) }.  \label{D010010 tilde}
\end{equation}%
Substituting here the appropriate term from Eq.~(\ref{phi (t tau)}) gives
\begin{equation}
\tilde{D}_{000,010}\left( t,t^{\prime }\right) =e^{i\omega \left( t^{\prime
}-t\right) -\gamma _{100}\left( t^{\prime }-t\right) -2\gamma _{010}t}%
\overline{C_{110}^{\ast }\left( t\right) C_{110}\left( t^{\prime }\right) }.
\label{D010010 tilde '}
\end{equation}

Substituting Eqs. (\ref{d 010010}) and (\ref{D010010 tilde '}) into Eq.~(\ref%
{K t, tau}), we arrive at
\begin{eqnarray}
K\left( t,\tau \right) &=&e^{-i\omega \tau -\left( \gamma _{100}+\frac{%
\Gamma }{2}\right) \tau -\Gamma t}\sin \left( \tilde{\Omega}_{R}t\right)
\sin \left[ \tilde{\Omega}_{R}\left( t+\tau \right) \right]  \notag \\
&&+e^{-i\omega \tau -\gamma _{010}\tau }\mu _{\Omega }
\left[ \frac{2\tilde{\Omega}_{R}^{2}}{\gamma _{n}(4\tilde{\Omega}%
_{R}^{2}+\gamma _{n}^{2})}e^{-2\gamma _{010}t}-\frac{1}{2\gamma _{n}}%
e^{-\Gamma t} \right.  \notag \\
&& \left. 
- \frac{1}{4\left( 2i\tilde{\Omega}_{R}-\gamma _{n}\right) }e^{2i%
\tilde{\Omega}_{R}t-\Gamma t}+\frac{1}{4\left( 2i\tilde{\Omega}_{R}+\gamma
_{n}\right) }e^{-2i\tilde{\Omega}_{R}t-\Gamma t}\right]  \notag \\
&&+e^{-i\omega \tau -\gamma _{010}\tau -\Gamma t}\frac{\mu _{\Omega }}{4}%
\left[ \frac{e^{\left( -\gamma _{d}+i\tilde{\Omega}_{R}\right) \tau }-1}{%
-\gamma _{d}+i\tilde{\Omega}_{R}}\left( 1-e^{2i\tilde{\Omega}_{R}t}\right)
+c.c.\right] .  \label{K(t,tau) final}
\end{eqnarray}%
where we denoted $\Gamma =\gamma _{110}+\gamma _{001}$, $\gamma _{n}=$ $%
\Gamma -2\gamma _{010}$, $\gamma _{d}=\gamma _{100}+\frac{\Gamma }{2}-\gamma
_{010}$.

Now we have everything to determine the emission spectra given by Eq.~(\ref%
{S(v)}). Using the values of $\gamma _{100}=\frac{\mu _{\Omega }}{2}$ and $%
\gamma _{010}=\frac{\mu _{\omega }}{2}$, we obtain
\begin{equation}
S\left( \nu \right) =\frac{1}{\pi }\mathrm{Re}\int_{0}^{\infty }d\tau
e^{i\nu \tau }\int_{0}^{\infty }dtK\left( t,\tau \right) =S_{1}\left(
\nu \right) +S_{2}\left( \nu \right) +S_{3}\left( \nu \right)
,  \label{es}
\end{equation}%
where%
\begin{eqnarray}
S_{1}\left( \nu \right) &=&\frac{2\tilde{\Omega}_{R}^{2}}{\pi \Gamma (4%
\tilde{\Omega}_{R}^{2}+\Gamma ^{2})}\mathrm{Re}\frac{\Gamma +\frac{\mu
_{\Omega }}{2}-i\left( \nu -\omega \right) }{[\gamma _{\mathrm{ac}%
}-i\left( \nu -\omega \right)]^{2}+\tilde{\Omega}_{R}^{2}}  \notag \\
S_{2}\left( \nu \right) &=&\frac{\tilde{\Omega}_{R}^{2}}{\pi \Gamma (4%
\tilde{\Omega}_{R}^{2}+\Gamma ^{2})}\frac{\mu _{\Omega }}{\frac{\mu _{\omega
}^{2}}{4}+\left( \nu -\omega \right) ^{2}}  \label{emission spectra} \\
S_{3}\left( \nu \right) &=&\frac{\mu _{\Omega }\tilde{\Omega}_{R}^{2}}{%
\pi \Gamma \left( \gamma _{d}^{2}+\tilde{\Omega}_{R}^{2}\right) (4\tilde{%
\Omega}_{R}^{2}+\Gamma ^{2})}\times  \notag \\
&&\left\{ -\mathrm{Re}\frac{\Gamma _{d}\left[ \gamma _{\mathrm{ac}}-i\left(
\nu -\omega \right) \right] +2\tilde{\Omega}_{R}^{2}-\gamma _{d}\Gamma }{%
\left[ \gamma _{\mathrm{ac}}-i\left( \nu -\omega \right) \right] ^{2}+\tilde{%
\Omega}_{R}^{2}}+\frac{\Gamma _{d}\frac{\mu _{\omega }}{2}}{\frac{\mu
_{\omega }^{2}}{4}+\left( \nu -\omega \right) ^{2}}\right\}  \notag
\end{eqnarray}%
The parameters $\Gamma $, $\gamma _{\mathrm{ac}}$, $\gamma _{d}$, and $%
\Gamma _{d}$ are expressed through the relaxation rates of the electron,
photon, and phonon subsystems $\mu _{\omega }$, $\mu _{\Omega }$ and $\gamma
$ as
\begin{eqnarray*}
\Gamma &=&\gamma _{110}+\gamma _{001}=\frac{1}{2}\left( \mu _{\omega }+\mu
_{\Omega }+\gamma \right) ,\ \gamma _{\mathrm{ac}}=\gamma _{100}+\frac{%
\Gamma }{2}=\frac{1}{4}\left( \mu _{\omega }+\gamma \right) +\frac{3}{4}\mu
_{\Omega }, \\
\gamma _{d} &=&\gamma _{100}+\frac{\Gamma }{2}-\gamma _{010}=\frac{1}{4}%
\left( \gamma -\mu _{\omega }\right) +\frac{3}{4}\mu _{\Omega },\ \Gamma
_{d}=\Gamma +2\gamma _{d}=2\mu _{\Omega }+\gamma .
\end{eqnarray*}%
The expression for the power spectrum $S\left( \nu \right) $ contains
the terms $S_{2}$ and $S_{3}$ which originate from the correlators $%
D_{010,010}$ and $\tilde{D}_{000,010}$ respectively. The term $S_{3}\left(
\nu \right) $ consists of two terms which have the same spectral shapes
as the functions $S_{1}\left( \nu \right) $ and $S_{2}\left( \nu
\right) $ respectively. Therefore, including the term $S_{3}\left( \nu
\right) $ in Eq.~(\ref{es}) leads only to corrections to the amplitudes of
the functions $S_{1,2}\left( \nu \right) $; moreover, under the strong
coupling conditions $\tilde{\Omega}_{R}\gg $ $\gamma _{ani}$ these
corrections are small: of the order of $\sim \frac{\mu _{\Omega }\left(
\Gamma +\gamma _{\mathrm{ac}}\right) }{\tilde{\Omega}_{R}^{2}}$ for the
function $S_{1}\left( \nu \right) $ and of the order of $\sim \frac{%
\Gamma _{d}\mu _{\omega }}{\tilde{\Omega}_{R}^{2}}$ for the function $%
S_{2}\left( \nu \right) $. For qualitative discussion we will neglect the
contribution of $S_{3}\left( \nu \right) $ and keep only the terms $%
S_{1}\left( \nu \right) $ and $S_{2}\left( \nu \right) $, although all terms are included in the spectra plotted in Fig.~4.

%%%%%%%%%%%%%%%%%%%%%%%%%%%%%%%%%%%%%

\begin{figure}[htb]
\includegraphics[width=0.6\linewidth]{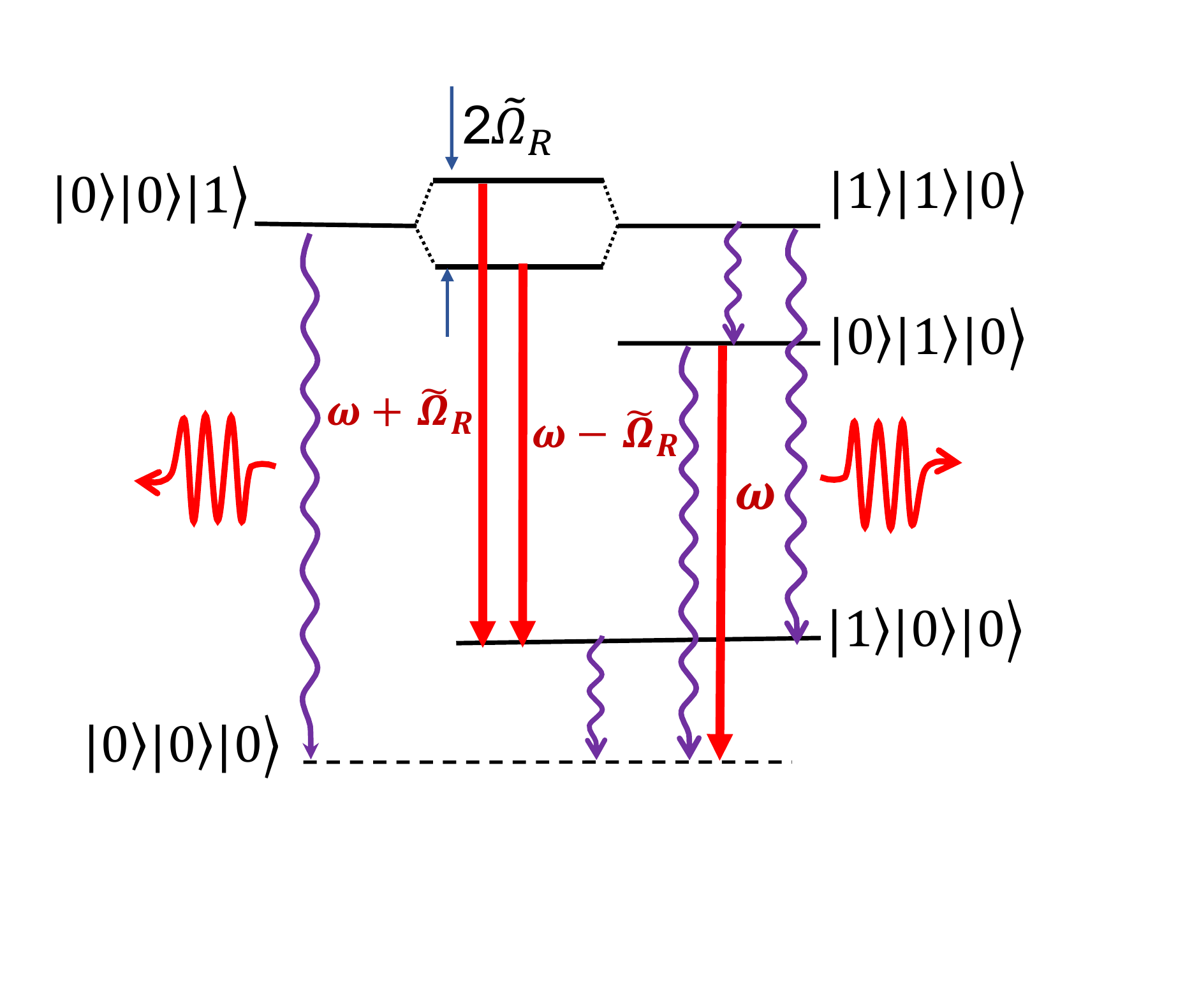}
\caption{ Energy levels of $ \left\vert phonon\right\rangle \left\vert photon\right\rangle \left\vert electron\right\rangle $ states  involved into the photon emission in the parametric decay of a single-electron excitation in a coupled phonon-photon-electron system. Bold red arrows indicate photon emission transitions with their peak frequencies labeled. Wavy purple arrows indicate various relaxation pathways.  }
\label{fig3}
\end{figure}

\begin{figure}[htb]
\includegraphics[width=0.5\linewidth]{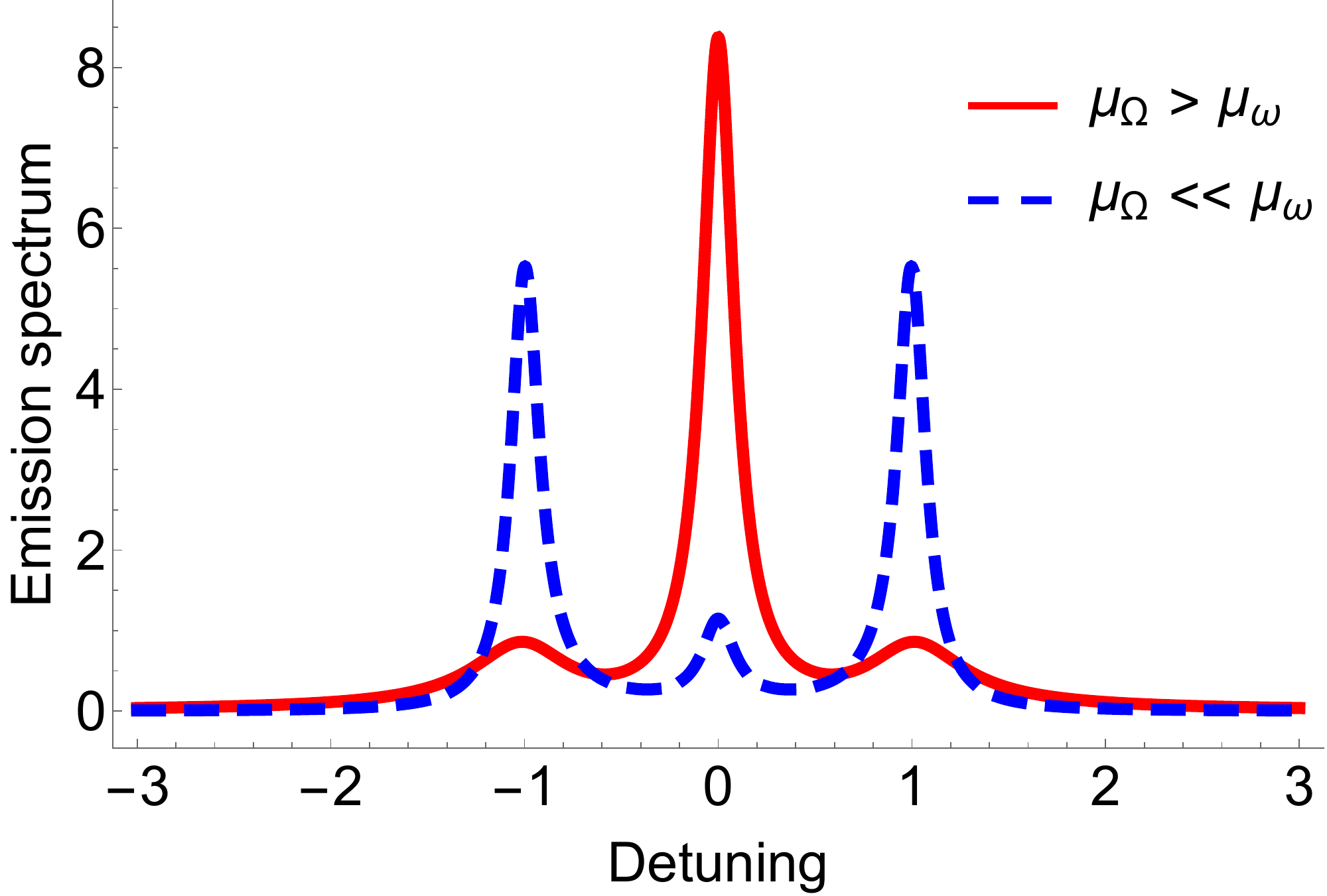}
\caption{ Normalized photon emission spectra $ \tilde{\Omega}_{R}^2 S(\nu) $ as a function of normalized detuning $\frac{\nu - \omega}{\tilde{\Omega}_{R}}$ from the cavity mode frequency $\omega$, for two different values of the phonon relaxation rate: $\mu_{\Omega} = 0.3$ (red solid line) and $\mu_{\Omega} = 0.02$ (blue dashed line).  Other relaxation rates are $\mu_{\omega} = 0.2$ and $\gamma = 0.1$. All relaxation constants are in units of $\tilde{\Omega}_{R}$.   }
\label{fig4}
\end{figure}

%%%%%%%%%%%%%%%%%%%%%%%%%

Figure 3 indicates all transitions that give contributions to the photon emission. 
The function $S_{1}\left( \nu \right) $ describes the emission spectrum
at the transition $\left\vert 1\right\rangle \left\vert 1\right\rangle
\left\vert 0\right\rangle \rightarrow \left\vert 1\right\rangle \left\vert
0\right\rangle \left\vert 0\right\rangle $, which is split due to Rabi
oscillations. The width of the peaks located at frequencies $\nu =\omega \pm
\tilde{\Omega}_{R}$ is equal to $\gamma _{\mathrm{ac}}=\frac{\mu _{\Omega }}{%
2}+\frac{\Gamma }{2}$. Here $\frac{\Gamma }{2}=\frac{1}{4}\left( \mu
_{\omega }+\mu _{\Omega }+\gamma \right) =\gamma _{MIX}$ is the decay rate
of the entangled state $\left\vert MIX\right\rangle =A\left( t\right)
\left\vert 0\right\rangle \left\vert 0\right\rangle \left\vert
1\right\rangle +B\left( t\right) \left\vert 1\right\rangle \left\vert
1\right\rangle \left\vert 0\right\rangle $ (see Eq. (\ref{solution for the
5-component state vector})), and $\frac{\mu _{\Omega }}{2}$ is the
broadening of the state $\left\vert 1\right\rangle \left\vert 0\right\rangle
\left\vert 0\right\rangle $ due to relaxation. The spectrum given by $%
S_{1}\left( \nu \right) $ agrees with the one obtained in \cite{tokman2021}.

The function $S_{2}\left( \nu \right) $ describes the emission due to a
two-step relaxation process described in section IV.C: $\left\vert
1\right\rangle \left\vert 1\right\rangle \left\vert 0\right\rangle
\rightarrow \left\vert 0\right\rangle \left\vert 1\right\rangle \left\vert
0\right\rangle \rightarrow \left\vert 0\right\rangle \left\vert
0\right\rangle \left\vert 0\right\rangle $ . The photons are emitted at the
transition $\left\vert 0\right\rangle \left\vert 1\right\rangle \left\vert
0\right\rangle \rightarrow \left\vert 0\right\rangle \left\vert
0\right\rangle \left\vert 0\right\rangle $, which is not affected by Rabi
oscillations; see Fig.~3. Therefore, this contribution has a standard Lorentzian shape
of an emitter at frequency $\omega $:%
\begin{equation*}
S_{2}\left( \nu \right) \propto \frac{1}{\gamma _{010}^{2}+\left( \nu
-\omega \right) ^{2}}=\frac{1}{\frac{\mu _{\omega }^{2}}{4}+\left( \nu
-\omega \right) ^{2}}.
\end{equation*}%
In the strong-coupling regime $\tilde{\Omega}_{R}\gg $ $\gamma _{ani}$ the
ratio of the amplitude of this central peak at frequency $\nu =\omega $ to
the amplitudes of the split peaks at frequencies $\nu =\omega \pm \tilde{%
\Omega}_{R}$ is given by
\begin{equation}
\frac{S_{2}\left( \omega \right) }{S_{1}\left( \omega \pm \tilde{\Omega}%
_{R}\right) }\approx \frac{\mu _{\Omega }\left( \mu _{\omega }+\gamma +3\mu
_{\Omega }\right) }{\mu _{\omega }^{2}}.
\label{the amplitudes of the split peaks}
\end{equation}

When $\mu _{\Omega }\rightarrow 0$, Eq.~(\ref{the amplitudes of the split
peaks}) gives $\frac{S_{2}\left( \omega \right) }{S_{1}\left( \omega \pm
\tilde{\Omega}_{R}\right) }\rightarrow 0$: indeed without phonon relaxation
the state $\left\vert 0\right\rangle \left\vert 1\right\rangle \left\vert
0\right\rangle $ cannot be populated from $\left\vert 1\right\rangle
\left\vert 1\right\rangle \left\vert 0\right\rangle $; therefore, the
two-step radiation channel is suppressed. In the opposite limit of a fast
phonon relaxation, when $\mu _{\Omega }\gg $ $\mu _{\omega },\gamma $, we
obtain $\frac{S_{2}\left( \omega \right) }{S_{1}\left( \omega \pm \tilde{%
\Omega}_{R}\right) }\gg 1$, i.e., the side peaks are weaker and more
broadened than the central peak. The latter statement is true despite the
large Rabi frequency $\tilde{\Omega}_{R}\gg $ $\mu _{\Omega }$. Indeed, when
$\tilde{\Omega}_{R}\gg $ $\mu _{\Omega }$ the interaction has the time to
mix the states $\left\vert 1\right\rangle \left\vert 1\right\rangle
\left\vert 0\right\rangle $ and $\left\vert 0\right\rangle \left\vert
0\right\rangle \left\vert 1\right\rangle $ before the phonon relaxation
kicks in. Nevertheless, if $\mu _{\Omega }\gg $ $\mu _{\omega },\gamma $ the
phonon relaxation is able to transfer population to state $\left\vert
0\right\rangle \left\vert 1\right\rangle \left\vert 0\right\rangle $ faster
than the radiative transition $\left\vert 1\right\rangle \left\vert
1\right\rangle \left\vert 0\right\rangle \rightarrow \left\vert
1\right\rangle \left\vert 0\right\rangle \left\vert 0\right\rangle $
corresponding to the Rabi-split spectrum.

This behavior is illustrated in Fig.~4 which shows photon emission spectra given by Eqs.~(\ref{es}) and (\ref{emission spectra}) as a function of frequency detuning $\nu - \omega$ from the cavity mode resonance, for two different values of the phonon relaxation rate: $\mu_{\Omega} = 1.5 \mu_{\omega}$ (red solid line) and $\mu_{\Omega} = 0.1 \mu_{\omega}$ (blue dashed line).  The electron relaxation rate is kept at $\gamma = 0.1$ and its exact value is not important for the overall shape of the spectra, although it affects absolute values and widths of the peaks. All quantities are normalized by $\tilde{\Omega}_{R}$. 

The relative magnitudes of the peaks and their widths depend sensitively on different combinations of the relaxation rates $\gamma, \mu_{\omega}, \mu_{\Omega}$. The onset of the strong coupling regime in the frequency domain is determined by the visibility of nonlinear Rabi splitting between the side peaks in Fig.~4, i.e., the condition  $\tilde{\Omega}_{R} > \frac{\gamma_{ac}}{2}$. We point out again that the relaxation rates of the individual subsystems enter the quantum dynamics in a very nontrivial way, and one need to know all of them to evaluate the feasibility of strong coupling in any particular system. The reverse is also true: once the  strong coupling regime is reached, 
 measurements of the photoluminescence spectra yield both the relaxation rates and the nonlinear coupling strength in the system. 
 
 A more detailed discussion of the feasibility of strong coupling at the nonlinear resonance in particular systems can be found in \cite{tokman2021}. Here we only point out that in dielectric microcavities the photon relaxation rates can be very low, in the $\mu$eV range, and the strong coupling threshold is likely to be determined by relaxation of the electron or vibrational transitions. In plasmonic nanocavities the photon relaxation rate can easily be tens of meV and will likely dominate the strong coupling threshold. On the other hand, the nonlinear coupling strength $\Omega_R^{(3)}$ is much higher in plasmonic nanocavities because of greatly enhanced electric field localization and electric field gradient. One can obtain  the magnitude of $\Omega_R^{(3)}$ of the order of 100 meV for the field localization in the few nm range, which is now routinely demonstrated in plasmonic  nanocavities.

%%%%%%%%%%%%%%%%%%%%%%%%%

\subsection{The phonon emission spectra}

There is a complete symmetry for the two bosonic fields in the decay process
close to the parametric resonance $\omega_e=\omega +\Omega $. Therefore,
we can obtain the phonon emission spectrum from the expressions for the
photon emission spectrum Eqs.~(\ref{es}), (\ref{emission spectra}), after
replacing
\begin{equation*}
\omega \Longleftrightarrow \Omega ,\ \mu _{\omega }\Longleftrightarrow \mu
_{\Omega },\ \gamma _{010}\Longleftrightarrow \gamma _{100}.
\end{equation*}

This results in
\begin{equation}
S_{p}\left( \nu \right) =S_{1p}\left( \nu \right) +S_{2p}\left(
\nu \right) +S_{3p}\left( \nu \right) ,  \label{spv}
\end{equation}%
where%
\begin{eqnarray}
S_{1p}\left( \nu \right) &=&\frac{2\tilde{\Omega}_{R}^{2}}{\pi \Gamma (4%
\tilde{\Omega}_{R}^{2}+\Gamma ^{2})}\mathrm{Re}\frac{\Gamma +\frac{\mu
_{\Omega }}{2}-i\left( \nu -\Omega \right) }{[\tilde{\gamma}_{\mathrm{%
ac}}-i\left( \nu -\Omega \right)]^{2}+\tilde{\Omega}_{R}^{2}}  \notag
\\
S_{2p}\left( \nu \right) &=&\frac{\tilde{\Omega}_{R}^{2}}{\pi \Gamma (4%
\tilde{\Omega}_{R}^{2}+\Gamma ^{2})}\frac{\mu _{\omega }}{\frac{\mu _{\Omega
}^{2}}{4}+\left( \nu -\Omega \right) ^{2}}  \label{emission spectra p} \\
S_{3p}\left( \nu \right) &=&\frac{\mu _{\omega }\tilde{\Omega}_{R}^{2}}{%
\pi \Gamma \left( \tilde{\gamma}_{d}^{2}+\tilde{\Omega}_{R}^{2}\right) (4%
\tilde{\Omega}_{R}^{2}+\Gamma ^{2})}\times  \notag \\
&&\left\{ -\mathrm{Re}\frac{\tilde{\Gamma}_{d}\left[ \tilde{\gamma}_{\mathrm{%
ac}}-i\left( \nu -\Omega \right) \right] +2\tilde{\Omega}_{R}^{2}-\tilde{%
\gamma}_{d}\Gamma }{\left[ \tilde{\gamma}_{\mathrm{ac}}-i\left( \nu -\Omega
\right) \right] ^{2}+\tilde{\Omega}_{R}^{2}}+\frac{\tilde{\Gamma}_{d}\frac{%
\mu _{\Omega }}{2}}{\frac{\mu _{\Omega }^{2}}{4}+\left( \nu -\Omega \right)
^{2}}\right\}  \notag
\end{eqnarray}
\begin{eqnarray*}
\tilde{\gamma}_{\mathrm{ac}} &=&\gamma _{010}+\frac{\Gamma }{2}=\frac{1}{4}%
\left( \mu _{\Omega }+\gamma \right) +\frac{3}{4}\mu _{\omega },\ \tilde{%
\gamma}_{d}=\gamma _{010}+\frac{\Gamma }{2}-\gamma _{100}=\frac{1}{4}\left(
\gamma -\mu _{\Omega }\right) +\frac{3}{4}\mu _{\omega } \\
\ \tilde{\Gamma}_{d} &=&\Gamma +2\tilde{\gamma}_{d}=2\mu _{\omega }+\gamma .
\end{eqnarray*}%
Similarly to the photon spectrum, the term $S_{3p}\left( \nu \right) $
is the sum of two terms which have the same spectral shape as $S_{1p}\left(
\nu \right) $ and $S_{2p}\left( \nu \right) $, but much smaller
magnitudes if $\tilde{\Omega}_{R}\gg $ $\gamma _{ani}$. Therefore we will
again include only the first two terms in qualitative discussion, but include all terms when plotting the spectra. 

%%%%%%%%%%%%%%%%%%%%%%%%%%%%%%%%%%%%%

\begin{figure}[htb]
\includegraphics[width=0.6\linewidth]{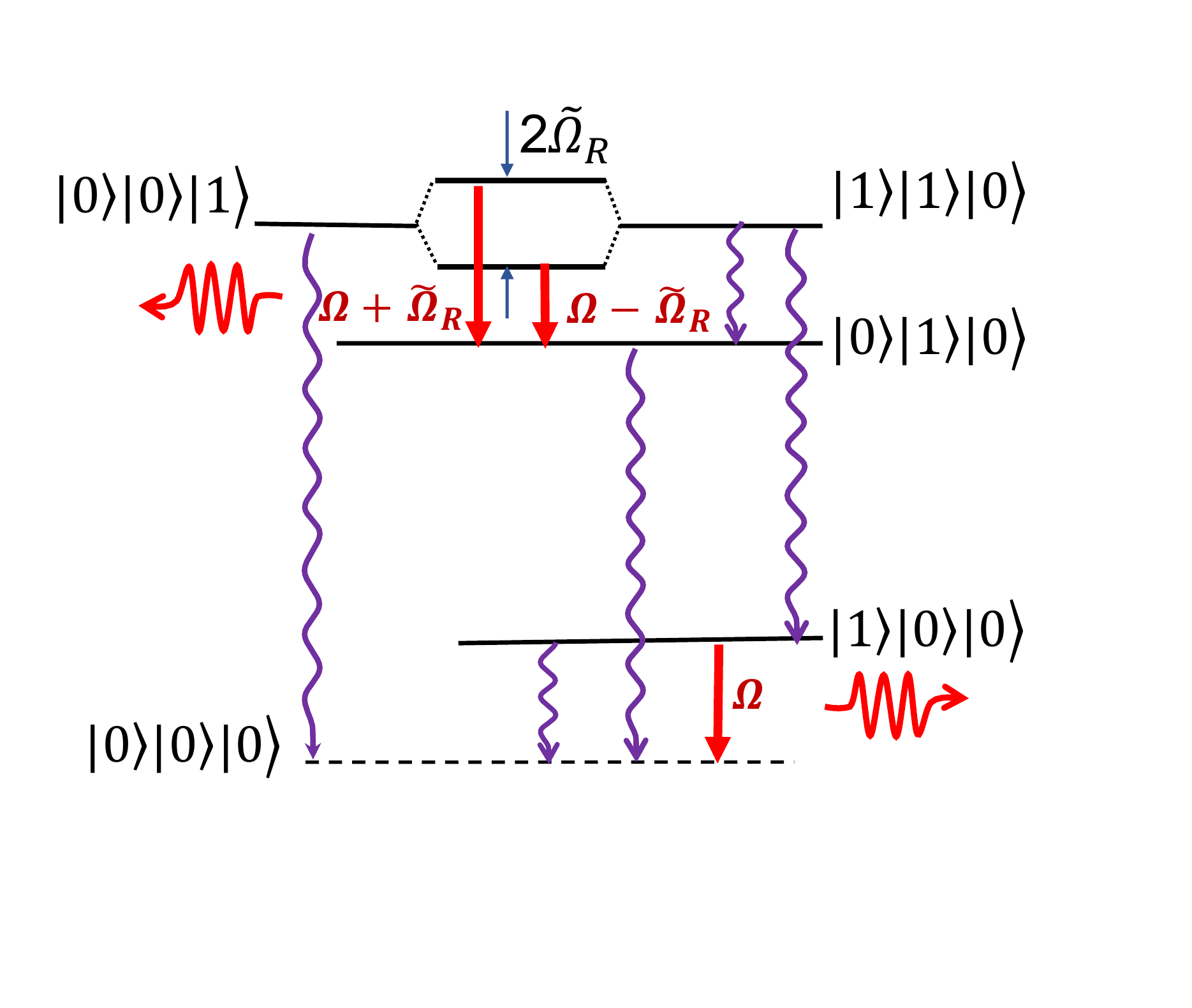}
\caption{ Energy levels of $ \left\vert phonon\right\rangle \left\vert photon\right\rangle \left\vert electron\right\rangle $ states involved into the phonon emission in the parametric decay of a single-electron excitation in a coupled phonon-photon-electron system. Bold red arrows indicate phonon emission transitions with their peak frequencies labeled. Wavy purple arrows indicate various relaxation pathways.  }
\label{fig5}
\end{figure}

\begin{figure}[htb]
\includegraphics[width=0.5\linewidth]{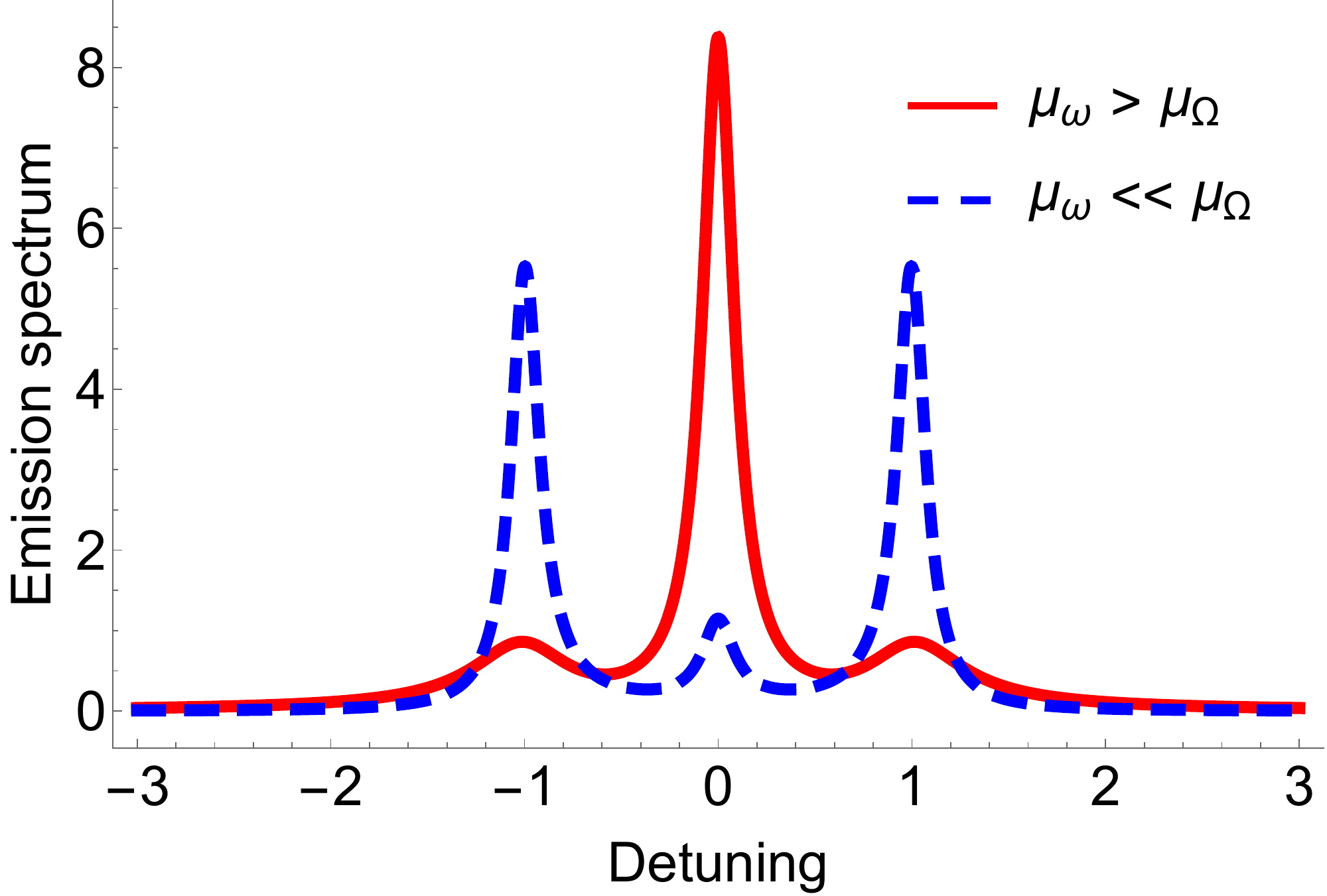}
\caption{ Normalized phonon emission spectra $\tilde{\Omega}_{R}^2 S_p(\nu)  $  as a function of normalized detuning $\frac{\nu - \Omega}{\tilde{\Omega}_{R}}$ from the vibrational mode frequency $\Omega$, for two different values of the photon relaxation rate: $\mu_{\omega} = 0.3$ (red solid line) and $\mu_{\omega} = 0.02$ (blue dashed line).  Other relaxation rates are $\mu_{\Omega} = 0.2$ and $\gamma = 0.1$. All relaxation constants are in units of $\tilde{\Omega}_{R}$.   }
\label{fig6}
\end{figure}

%%%%%%%%%%%%%%%%%%%%%%%%%

Figure 5 shows all transitions giving contributions to the phonon emission spectrum, with their peak frequencies indicated. 

The function $S_{1p}\left( \nu \right) $ describes the phonon emission
spectrum due to the transition $\left\vert 1\right\rangle \left\vert
1\right\rangle \left\vert 0\right\rangle \rightarrow \left\vert
0\right\rangle \left\vert 1\right\rangle \left\vert 0\right\rangle $, which
demonstrates Rabi splitting. The width of the peaks centered at frequencies $%
\nu =\Omega \pm \tilde{\Omega}_{R}$ is equal to $\tilde{\gamma}_{\mathrm{ac}%
}=\frac{\mu _{\omega }}{2}+\frac{\Gamma }{2}$. The function $S_{2p}\left(
\nu \right) $ describes phonon emission due to a two-step relaxation $%
\left\vert 1\right\rangle \left\vert 1\right\rangle \left\vert
0\right\rangle \rightarrow \left\vert 1\right\rangle \left\vert
0\right\rangle \left\vert 0\right\rangle \rightarrow \left\vert
0\right\rangle \left\vert 0\right\rangle \left\vert 0\right\rangle $. The
phonons are emitted at the second step, i.e. the transition $\left\vert
1\right\rangle \left\vert 0\right\rangle \left\vert 0\right\rangle
\rightarrow \left\vert 0\right\rangle \left\vert 0\right\rangle \left\vert
0\right\rangle $, which is not affected by Rabi oscillations. Therefore, the
spectrum due to this contribution is a standard Lorenzian line, similarly to
the case of photons:
\begin{equation*}
S_{2}\left( \nu \right) \propto \frac{1}{\gamma _{100}^{2}+\left( \nu
-\Omega \right) ^{2}}=\frac{1}{\frac{\mu _{\Omega }^{2}}{4}+\left( \nu
-\Omega \right) ^{2}}.
\end{equation*}%
The ratio of the amplitude of the central peak at $\nu =\Omega $ to those of
the side peaks at frequencies $\nu =\Omega \pm \tilde{\Omega}_{R}$ at $%
\tilde{\Omega}_{R}\gg $ $\gamma _{ani}$ is given by the expression
equivalent to Eq.~(\ref{the amplitudes of the split peaks}) after
substituting $\omega \Longleftrightarrow \Omega \ $and $\mu _{\omega
}\Longleftrightarrow \mu _{\Omega }$:
\begin{equation}
\frac{S_{2p}\left( \Omega \right) }{S_{1p}\left( \Omega \pm \tilde{\Omega}%
_{R}\right) }\approx \frac{\mu _{\omega }\left( \mu _{\Omega }+\gamma +3\mu
_{\omega }\right) }{\mu _{\Omega }^{2}}.  \label{aosp}
\end{equation}

The phonon emission spectra are plotted in Fig.~6 as a function of frequency detuning $\nu - \Omega$ from the cavity mode resonance. This time we keep the phonon relaxation rate fixed at $\mu_{\Omega} = 0.2$ and plot the spectra for two values of the photon relaxation rate, greater and smaller than $\mu_{\Omega}$:  $\mu_{\omega} = 0.3$ (red solid line) and $\mu_{\omega} = 0.02$ (blue dashed line).   All quantities are normalized by $\tilde{\Omega}_{R}$. The numbers are chosen to prove the point that the phonon and photon spectra are symmetric with respect to replacement indicated in the beginning of this section.

In experiment, measuring the ratios given by Eq.~(\ref{the amplitudes of the
split peaks}) and (\ref{aosp}) allows one to determine the relationships
between all relaxation rates $\mu _{\Omega }$, $\gamma $, and $\mu _{\omega
} $. Indeed, one can obtain from Eqs.~(\ref{the amplitudes of the split
peaks}) and (\ref{aosp}) that
\begin{equation*}
\xi _{\Omega }x^{3}+2x^{2}-2x-\xi _{\omega }=0,\ y=\xi _{\Omega }x^{2}-3-x;
\end{equation*}%
where $x=\frac{\mu _{\Omega }}{\mu _{\omega }}$, $y=$ $\frac{\gamma }{\mu
_{\omega }}$, $\xi _{\omega }=\frac{S_{2}\left( \omega \right) }{S_{1}\left(
\omega \pm \tilde{\Omega}_{R}\right) }$, $\xi _{\Omega }=\frac{S_{2p}\left(
\Omega \right) }{S_{1p}\left( \Omega \pm \tilde{\Omega}_{R}\right) }$.

%%%%%%%%%%%%%%%%%%%%%%%%%%%%%

\section{Conclusions} 

We developed a universal model of three-wave nonlinear resonance which is applicable to the systems with coupled electron, photon, and vibrational degrees of freedom, for example in cavity QED with molecular quantum emitters or quantum dots and in cavity optomechanics systems. We obtained the analytic solution for the nonperturbative quantum dynamics of such systems in the vicinity of the nonlinear resonance, taking into account dissipation and fluctuations for all degrees of freedom in Markov approximation.  We showed that the strong coupling regime can be reached at the nonlinear resonance when the nonlinear coupling strength exceeds a certain threshold which includes all relaxation rates. When all three degrees of freedom are quantized, strong coupling leads to the formation of tripartite entangled states. The presented solution can be used to derive the explicit analytic expression for any observable. As an example, we calculated photon and phonon emission spectra which have a characteristic three-peak form once the strong coupling is reached. We showed how the relative heights and widths of the peaks can be used to extract information about all relaxation rates in the system and the nonlinear coupling strength, or to establish the threshold for reaching the strong coupling regime. 

%%%%%%%%%%%%%%%%%%%%%%%%%%%%%%%%%%%%%%%%%%%%

\begin{acknowledgments}
This work has been supported in part by the Air Force Office for Scientific Research 
Grant No.~FA9550-21-1-0272, National Science Foundation Award No.~1936276, and Texas A\&M University through STRP, X-grant and T3-grant programs.
 M.T.~and M.E.~acknowledge the support by the Center of Excellence
``Center of Photonics'' funded by The Ministry of
Science and Higher Education of the Russian Federation,
Contract No.~075-15-2020-906.

\end{acknowledgments}

%%%%%%%%%%%%%%%%%%%%%%%%%%%%%

\appendix

\section{The derivation of Eqs.~(\protect\ref{c100 2})-(\protect\ref{c000 2})%
}

A number of correlators of the random functions in Eq. (\ref{solution for
the 5-component state vector}) is zero due to Eqs.~(\ref{diagonal elements}%
),(\ref{off-diagonal elements}):
\begin{equation}
\overline{\delta C_{110}^{\ast }\left( t^{\prime }\right) \mathfrak{R}%
_{\alpha mi}\left( t^{\prime \prime }\right) }=\overline{\delta
C_{001}^{\ast }\left( t^{\prime }\right) \mathfrak{R}_{\alpha mi}\left(
t^{\prime \prime }\right) }=0  \label{b1}
\end{equation}%
\begin{equation}
\overline{\delta C_{001}^{\ast }\left( t^{\prime }\right) \delta C_{001}\left(
t^{\prime \prime }\right) }=\overline{\delta C_{110}^{\ast }\left( t^{\prime
}\right) \delta C_{110}\left( t^{\prime \prime }\right) }=\overline{\delta
C_{001}^{\ast }\left( t^{\prime }\right) \delta C_{110}\left( t^{\prime \prime
}\right) }=0  \label{b2}
\end{equation}%
\begin{eqnarray}
\overline{\delta C_{001}^{\ast }\left( t^{\prime }\right) C_{000}\left(
t^{\prime \prime }\right) } &=&\overline{\delta C_{110}^{\ast }\left(
t^{\prime }\right) C_{100}\left( t^{\prime \prime }\right) }=\overline{%
\delta C_{001}^{\ast }\left( t^{\prime }\right) C_{010}\left( t^{\prime
\prime }\right) } 
=\overline{\delta C_{110}^{\ast }\left( t^{\prime }\right) C_{000}\left(
t^{\prime \prime }\right) }    \notag \\ 
&=& \overline{\delta C_{110}^{\ast }\left( t^{\prime
}\right) C_{100}\left( t^{\prime \prime }\right) }=\overline{\delta
C_{110}^{\ast }\left( t^{\prime }\right) C_{010}\left( t^{\prime \prime
}\right) } = 0.  \label{b3}
\end{eqnarray}%
Eqs.~(\ref{b1})-(\ref{b3}) ensure that the variables $\delta C_{100}$ and $%
\delta C_{110}$ cannot contribute to the values of any observables and
therefore can be omitted.

The other correlators are given by the equations that follow from Eqs. (\ref%
{solution1}):
\begin{equation*}
\frac{d}{dt}\overline{C_{100}^{\ast }C_{010}}=-\left( \gamma _{100}+\gamma
_{010}\right) \overline{C_{100}^{\ast }C_{010}}+D_{100,010},
\end{equation*}%
\begin{equation*}
\frac{d}{dt}\overline{C_{100}^{\ast }C_{000}}=-\gamma _{100}\overline{%
C_{100}^{\ast }C_{000}}+D_{100,000},
\end{equation*}%
\begin{equation*}
\frac{d}{dt}\overline{C_{010}^{\ast }C_{000}}=- \gamma _{010} \overline{C_{010}^{\ast }C_{000}}+D_{010,000},
\end{equation*}%
\begin{equation*}
\frac{d}{dt}\overline{\left\vert C_{000}\right\vert ^{2}}=D_{000,000},
\end{equation*}%
\begin{equation*}
\frac{d}{dt}\overline{\left\vert C_{010}\right\vert ^{2}}=-2\gamma _{010}%
\overline{\left\vert C_{010}\right\vert ^{2}}+D_{010,010},
\end{equation*}%
\begin{equation*}
\frac{d}{dt}\overline{\left\vert C_{100}\right\vert ^{2}}=-2\gamma _{100}%
\overline{\left\vert C_{100}\right\vert ^{2}}+D_{100,100}.
\end{equation*}%
Using Eqs.~(\ref{diagonal elements}),(\ref{off-diagonal elements}), we
arrive at

\begin{eqnarray}
\frac{d}{dt}\overline{C_{100}^{\ast }C_{010}} &=&-\frac{\mu _{\omega }+\mu
_{\Omega }}{2}\overline{C_{100}^{\ast }C_{010}}  \notag \\
\frac{d}{dt}\overline{C_{100}^{\ast }C_{000}} &=&-\frac{\mu _{\Omega }}{2}%
\overline{C_{100}^{\ast }C_{000}}+\mu _{\omega }\overline{C_{110}^{\ast
}C_{010}}  \label{b4} \\
\frac{d}{dt}\overline{C_{010}^{\ast }C_{000}} &=&-\frac{\mu _{\omega }}{2}%
\overline{C_{010}^{\ast }C_{000}}+\mu _{\Omega }\overline{C_{110}^{\ast
}C_{100}}  \notag
\end{eqnarray}

\bigskip
\begin{eqnarray}
\frac{d}{dt}\overline{\left\vert C_{000}\right\vert ^{2}} &=&\gamma
\overline{\left\vert C_{001}\right\vert ^{2}}+\mu _{\omega }\overline{%
\left\vert C_{010}\right\vert ^{2}}+\mu _{\Omega }\overline{\left\vert
C_{100}\right\vert ^{2}}  \notag \\
\frac{d}{dt}\overline{\left\vert C_{010}\right\vert ^{2}} &=&-\mu _{\omega }%
\overline{\left\vert C_{010}\right\vert ^{2}}+\mu _{\Omega }\overline{%
\left\vert C_{110}\right\vert ^{2}}  \label{b5} \\
\frac{d}{dt}\overline{\left\vert C_{100}\right\vert ^{2}} &=&-\mu _{\Omega }%
\overline{\left\vert C_{100}\right\vert ^{2}}+\mu _{\omega }\overline{%
\left\vert C_{110}\right\vert ^{2}}  \notag
\end{eqnarray}%
Taking into account Eqs.~(\ref{b3}), one can obtain that $\overline{%
C_{110}^{\ast }C_{010}}=\overline{C_{110}^{\ast }C_{100}}$ $=0$ in Eqs.~(\ref%
{b4}); as a result, for our initial conditions Eqs.~(\ref{b4}) yield
\begin{equation}
\overline{C_{100}^{\ast }C_{010}}=\overline{C_{100}^{\ast }C_{000}}=%
\overline{C_{010}^{\ast }C_{000}}=0.  \label{b6}
\end{equation}

The last two equations in Eqs.~(\ref{b5}) give
\begin{equation*}
\overline{\left\vert C_{010}\right\vert ^{2}}=\mu _{\Omega }e^{-\mu _{\omega
}t}\int_{0}^{t}e^{\mu _{\omega }\tau }\overline{\left\vert
C_{110}\right\vert ^{2}}d\tau ,\ \ \overline{\left\vert C_{100}\right\vert
^{2}}=\mu _{\omega }e^{-\mu _{\Omega }t}\int_{0}^{t}e^{\mu _{\Omega }\tau }%
\overline{\left\vert C_{110}\right\vert ^{2}}d\tau .
\end{equation*}%
Substituting here the function from Eq. (\ref{solution2}) and taking into
account Eqs.~(\ref{b2}),(\ref{b3}) results in
\begin{equation*}
\overline{\left\vert C_{100}\right\vert ^{2}}=\mu _{\omega }e^{-\mu _{\Omega
}t}\int_{0}^{t}e^{\frac{\mu _{\Omega }-\mu _{\omega }-\gamma }{2}\tau }\sin
^{2}\left( \tilde{\Omega}_{R}\tau \right) d\tau ,\ \ \overline{\left\vert
C_{010}\right\vert ^{2}}=\mu _{\Omega }e^{-\mu _{\omega }t}\int_{0}^{t}e^{%
\frac{\mu _{\omega }-\mu _{\Omega }-\gamma }{2}\tau }\sin ^{2}\left( \tilde{%
\Omega}_{R}\tau \right) d\tau ;
\end{equation*}%
For further integration we use the limit $\frac{\gamma _{\alpha ni}}{\tilde{%
\Omega}_{R}}\ll 1$, leading to
\begin{equation}
\overline{\left\vert C_{100}\right\vert ^{2}}\approx \frac{\mu _{\omega }}{%
\mu _{\Omega }-\mu _{\omega }-\gamma }\left( e^{-\frac{\mu _{\Omega }+\mu
_{\omega }+\gamma }{2}t}-e^{-\mu _{\Omega }t}\right)  \label{b7}
\end{equation}%
\begin{equation}
\overline{\left\vert C_{010}\right\vert ^{2}}\approx \frac{\mu _{\Omega }}{%
\mu _{\omega }-\mu _{\Omega }-\gamma }\left( e^{-\frac{\mu _{\Omega }+\mu
_{\omega }+\gamma }{2}t}-e^{-\mu _{\omega }t}\right)  \label{b8}
\end{equation}%
Note that Eqs.~(\ref{b7}),(\ref{b8}) do not contain any divergence when $%
\left[ \pm \left( \mu _{\omega }-\mu _{\Omega }\right) -\gamma \right]
\longrightarrow 0$. Indeed,
\begin{equation*}
\lim_{\left( \mu _{\Omega }-\mu _{\omega }-\gamma \right) \longrightarrow 0}
\left[ \frac{e^{-\frac{\mu _{\Omega }+\mu _{\omega }+\gamma }{2}t}-e^{-\mu
_{\Omega }t}}{\mu _{\Omega }-\mu _{\omega }-\gamma }\right] =\frac{1}{2}%
te^{-\mu _{\Omega }t},\ \lim_{\left( \mu _{\omega }-\mu _{\Omega }-\gamma
\right) \longrightarrow 0}\left[ \frac{e^{-\frac{\mu _{\Omega }+\mu _{\omega
}+\gamma }{2}t}-e^{-\mu _{\omega }t}}{\mu _{\omega }-\mu _{\Omega }-\gamma }%
\right] =\frac{1}{2}te^{-\mu _{\omega }t}.
\end{equation*}

Now we return to the first of Eqs.~(\ref{b5}), which yields
\begin{equation*}
\overline{\left\vert C_{000}\right\vert ^{2}}=\int_{0}^{t}\left( \gamma
\overline{\left\vert C_{001}\right\vert ^{2}}+\mu _{\omega }\overline{%
\left\vert C_{010}\right\vert ^{2}}+\mu _{\Omega }\overline{\left\vert
C_{100}\right\vert ^{2}}\right) d\tau .
\end{equation*}%
We substitute Eqs.~(\ref{b7}) and (\ref{b8}) into the second and third
terms in the integrand and substitute the expression $\overline{\left\vert
C_{001}\right\vert ^{2}}$ which follows from Eqs. (\ref{solution2}) into the
first term in the integrand. Neglecting the small terms $\propto $ $\frac{%
\gamma _{110}-\gamma _{001}}{\tilde{\Omega}_{R}}$and $\frac{\gamma _{\alpha
ni}}{\tilde{\Omega}_{R}}$ the integration results in
\begin{equation}
\overline{\left\vert C_{000}\right\vert ^{2}}=\frac{\gamma \left( \gamma
-\mu _{\Omega }-\mu _{\omega }\right) }{\gamma ^{2}-\left( \mu _{\Omega
}-\mu _{\omega }\right) ^{2}}\left( 1-e^{-\frac{\mu _{\Omega }+\mu _{\omega
}+\gamma }{2}t}\right) -\left( \mu _{\Omega }\frac{1-e^{-\mu _{\omega }t}}{%
\mu _{\omega }-\mu _{\Omega }-\gamma }+\mu _{\omega }\frac{1-e^{-\mu
_{\Omega }t}}{\mu _{\Omega }-\mu _{\omega }-\gamma }\right)  \label{b9}
\end{equation}

%%%%%%%%%%%%%%%%%%%%%%%%%%%%%

\end{document}